\documentclass[10pt,aps,prd,twocolumn,showpacs,superscriptaddress,nofootinbib,nobibnotes,longbibliography,floatfix]{revtex4-1}

\usepackage[utf8]{inputenc}
\usepackage[T1]{fontenc}

\usepackage{bm}
\usepackage{mathtools,amsmath,amssymb,amsfonts,mathrsfs,eucal,graphicx,tensor,csquotes,accents,commath,chngcntr,siunitx}
\usepackage[dvipsnames]{xcolor}
\usepackage[unicode]{hyperref}
\hypersetup{colorlinks=true, citecolor=MidnightBlue,
            linkcolor=MidnightBlue, urlcolor=MidnightBlue, linktocpage=true}
\usepackage[normalem]{ulem}
\usepackage{orcidlink}

\DeclareMathAlphabet{\mathpzc}{OT1}{pzc}{m}{it}
\definecolor{darkgreen}{rgb}{0.0, 0.6, 0.0}

\newcommand{\note}[1]{\text{\scshape\tiny{#1}}}

\newcommand{\ii}{\mathrm{i}}
\newcommand{\dd}{\mathrm{d}}

\newcommand{\rh}{r_\mathrm{H}}
\newcommand{\rM}{\bar{r}}

\newcommand{\al}{\alpha}

\newcommand{\de}{\delta}

\newcommand{\ze}{\zeta}

\renewcommand\th{\theta}

\newcommand{\La}{\Lambda}

\newcommand{\om}{\omega}

\newcommand{\orcid}[1]{\href{https://orcid.org/#1}{\includegraphics[width=10pt]{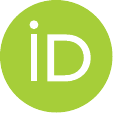}}}

\begin{document}

\author{Sebastian H. V\"olkel \orcid{0000-0002-9432-7690}}
\affiliation{SISSA, Via Bonomea 265, 34136 Trieste, Italy and INFN Sezione di Trieste}
\affiliation{IFPU - Institute for Fundamental Physics of the Universe, Via Beirut 2, 34014 Trieste, Italy}
\affiliation{Max Planck Institute for Gravitational Physics (Albert Einstein Institute), D-14476 Potsdam, Germany}
\email{sebastian.voelkel@aei.mpg.de}

\author{Nicola Franchini \orcid{0000-0002-9939-733X}}
\affiliation{SISSA, Via Bonomea 265, 34136 Trieste, Italy and INFN Sezione di Trieste}
\affiliation{IFPU - Institute for Fundamental Physics of the Universe, Via Beirut 2, 34014 Trieste, Italy}
\affiliation{Université Paris Cit\'e, CNRS, Astroparticule et Cosmologie,  F-75013 Paris, France}
\affiliation{CNRS-UCB International Research Laboratory, Centre Pierre Binétruy, IRL2007, CPB-IN2P3, Berkeley, US}

\author{Enrico Barausse \orcid{0000-0001-6499-6263}}
\affiliation{SISSA, Via Bonomea 265, 34136 Trieste, Italy and INFN Sezione di Trieste}
\affiliation{IFPU - Institute for Fundamental Physics of the Universe, Via Beirut 2, 34014 Trieste, Italy}

\author{Emanuele Berti \orcid{0000-0003-0751-5130}}
\affiliation{William H. Miller III Department of Physics and Astronomy, Johns Hopkins University, 3400 North Charles Street, Baltimore, Maryland, 21218, USA}

\date{\today}

\title{Constraining modifications of black hole perturbation potentials near the light ring with quasinormal modes}

\begin{abstract}
In modified theories of gravity, the potentials appearing in the Schr\"odinger-like equations that describe perturbations of non-rotating black holes are also modified. 
In this paper we ask: can these modifications be constrained with high-precision gravitational-wave measurements of the black hole's quasinormal mode frequencies? 
We  expand the modifications in a small perturbative parameter regulating the deviation from the general-relativistic potential, and in powers of $M/r$. 
We compute the quasinormal modes of the modified potential up to quadratic order in the perturbative parameter. 
Then we use Markov-chain-Monte-Carlo (MCMC) methods to recover the coefficients in the $M/r$ expansion in an ``optimistic'' scenario where we vary them one at a time, and in a ``pessimistic'' scenario where we vary them all simultaneously. 
In both cases, we find that the bounds on the individual parameters are not robust. 
Because quasinormal mode frequencies are related to the behavior of the perturbation potential near the light ring, we propose a different strategy. 
Inspired by Wentzel–Kramers–Brillouin (WKB) theory, we demonstrate that the value of the potential and of its second derivative at the light ring can be robustly constrained. 
These constraints allow for a more direct comparison between tests based on black hole spectroscopy and observations of black hole ``shadows'' by the Event Horizon Telescope and future instruments. 
\end{abstract}

\maketitle

\section{Introduction}

Since the first groundbreaking direct detection of gravitational waves in 2015~\cite{LIGOScientific:2016aoc}, the LIGO-Virgo-KAGRA collaboration has reported around one hundred additional events~\cite{LIGOScientific:2018mvr,LIGOScientific:2020ibl,LIGOScientific:2021djp,Nitz:2021zwj,Olsen:2022pin}. 
Most of the detected signals are in agreement with the predictions of general relativity (GR) for the merger of two black holes, while some of them involve neutron stars. 
Besides the important implications for astrophysical formation scenarios of black hole populations, these events allow for unprecedented tests of GR in the strong field~\cite{Berti:2015itd,Barack:2018yly,Yunes:2016jcc,Cardoso:2016ryw,Berti:2018cxi,Berti:2018vdi,Cardoso:2019rvt,LIGOScientific:2016lio,LIGOScientific:2019fpa,LIGOScientific:2020tif,LIGOScientific:2021sio,Ghosh:2021mrv}. 
A key prediction from GR is that rotating astrophysical black holes should be described by the Kerr metric, and that their perturbative dynamics at late times (in the so-called ``ringdown'' regime) should be well described by a superposition of damped sinusoids called quasinormal modes (QNMs), with characteristic frequencies determined only by the black hole's mass and spin. 
If GR is the correct theory of gravity, the observed QNM frequencies should match those predicted in GR for Kerr black holes~\cite{Detweiler:1980gk,Dreyer:2003bv,Berti:2005ys}. 

The LIGO-Virgo-KAGRA collaboration has analyzed the current catalogs, at first focusing on GW150914 and then using multiple events, with the specific aim to extract QNMs from the ringdown~\cite{LIGOScientific:2016lio,LIGOScientific:2019fpa,LIGOScientific:2020tif,LIGOScientific:2021sio}. 
The observations are generally consistent with the presence of the quadrupole ($\ell=m=2$) fundamental mode (with ``overtone number'' $n=0$). 
Reference~\cite{Isi:2019aib} claimed evidence for the presence of overtones (higher-damping $\ell=m=2$ modes with $n > 0$) in GW150914, and Ref.~\cite{Capano:2021etf} claimed evidence for the fundamental mode with $\ell=m=3$ in GW190521. 
Later work showed that these conclusions depend on the detector noise, data analysis methods, the choice of starting time, and nonlinear effects in the theoretical modeling of ringdown~\cite{Cotesta:2022pci,Finch:2022ynt,Isi:2022mhy,Capano:2022zqm,Sberna:2021eui,Cheung:2022rbm,Mitman:2022qdl}. 
Despite the ongoing debate, the analysis of numerical relativity simulations implies that at least some overtones can in principle be extracted when the signal-to-noise ratio is large enough~\cite{Berti:2007zu,Baibhav:2017jhs,Giesler:2019uxc,Ota:2019bzl,Bhagwat:2019dtm,Bhagwat:2019bwv,Cook:2020otn,Bustillo:2020buq,Forteza:2020hbw,MaganaZertuche:2021syq}. 
Theoretical modeling and data analysis challenges may be more subtle than anticipated, but next-generation detectors are expected to provide reliable, high-precision measurements of more than one QNM~\cite{Berti:2016lat,Cabero:2019zyt,Ota:2021ypb,Bhagwat:2021kwv}. 
Therefore it is important to investigate how these measurements will inform us on possible deviations from GR.

There have been various attempts to introduce deviations from GR in the QNM spectrum in a theory-agnostic manner. 
These include modifications of the gravitational action around a GR black hole background~\cite{Tattersall:2017erk}, the addition of perturbative corrections at the level of the perturbation equations~\cite{Cardoso:2019mqo,McManus:2019ulj,Volkel:2022aca} or at the level of the metric~\cite{Volkel:2020daa,Konoplya:2022pbc} (which then require suitable assumptions for the dynamics), or the addition of free parameters in the QNM frequencies themselves~\cite{Meidam:2014jpa,Maselli:2019mjd,Carullo:2021dui}. 
While the physical interpretation of these ``theory-agnostic'' constraints requires a specific modified theory of gravity, it is desirable to have a robust phenomenological framework encompassing several theories and allowing us to solve the ``inverse problem'' -- {\it i.e.}, to infer what specific GR modification caused a deviation from GR. 

For non-rotating spacetimes, the inverse problem based on parametrized black hole metrics for axial (odd parity) perturbations was studied in Ref.~\cite{Volkel:2020daa}. 
For spinning black holes, a parametrized spectroscopy framework (``ParSpec'') was introduced in Ref.~\cite{Maselli:2019mjd}, and applied to data in Ref.~\cite{Carullo:2021dui}. 
The ParSpec framework is based on a Taylor-series expansion of the QNM frequencies in the dimensionless Kerr spin parameter. 
In principle this can be used to stack multiple events~\cite{Yang:2017zxs}, and then compare with the predictions from specific theories.

In this work we assume that deviations from the QNM spectrum in GR can be adequately captured by small modifications of the underlying perturbation equations. 
As a proof of principle, we focus on non-rotating or slowly rotating black holes. 
We adopt the parametrized formalism of Refs.~\cite{Cardoso:2019mqo,McManus:2019ulj}, which systematically connects small deviations in the perturbation equations with the QNM spectrum. 
At lowest order, the idea is to consider small modifications in the perturbation potential at the linear level and to write them as a ``post-Newtonian'' (PN) series expansion in $M/r$~\cite{Cardoso:2019mqo}. 
Going to quadratic order allows one to capture (more realistically) possible couplings to additional fields~\cite{McManus:2019ulj}. 
While in Refs.~\cite{Cardoso:2019mqo,McManus:2019ulj} the focus is on the fundamental mode, Ref.~\cite{Volkel:2022aca} extended the analysis to overtones, and studied the inverse problem using a principal component analysis.

We apply a Bayesian analysis to solve the inverse problem: given a simulated set of QNM frequencies computed within the parametrized formalism, can we infer the deviation parameters in the underlying potentials? 
We are particularly interested in the understanding whether it is possible to constrain the more general and realistic case where many deviation parameters are varied simultaneously. 
Therefore we compute constraints on the individual PN-like expansion parameters twice: we first vary the parameters one at a time (``optimistic'' case), and then we vary them all simultaneously (``pessimistic'' case). 
In both cases we find that the constraints on the individual expansion parameters are not robust. 
The priors play an important role, especially when all parameters are varied simultaneously. 
In particular in the ``optimistic'' case the recovered bounds on the individual parameters can be biased. 

However, the complex correlations in the posterior distributions in the ``pessimistic'' case contain valuable information. 
It has long been known that QNM frequencies are related to the behavior of the perturbation potential near the light ring~\cite{Press:1971wr,Schutz:1985km,Cardoso:2008bp}. 
Therefore we propose a different strategy: we map the PN series expansion to the value of the potential and its derivatives at the peak by using insights from Wentzel–Kramers–Brillouin (WKB) theory. 
Higher-order WKB methods relate the QNM frequencies with a Taylor expansion of the effective perturbation potential in the vicinity of its maximum, i.e., close to the light-ring~\cite{Schutz:1985km,Iyer:1986np,Konoplya:2003ii,Matyjasek:2017psv}. 
In fact, the (closely related) eikonal approximation was used in Refs.~\cite{Glampedakis:2017dvb,Glampedakis:2019dqh,Silva:2019scu,Bryant:2021xdh} to build a  ``post-Kerr'' strategy to parametrized black hole spectroscopy. 
By connecting the Taylor (PN-like) expansion to the light-ring WKB expansion, we demonstrate that the value of the potential and of its second derivative at the light ring can be robustly constrained using Bayesian techniques.
 
Our work suggests that it may be possible to relate black hole spectroscopy tests to electromagnetic observations of black hole ``shadows'' by the Event Horizon Telescope and future instruments~\cite{EventHorizonTelescope:2019dse,EventHorizonTelescope:2019ggy,Volkel:2020xlc,EventHorizonTelescope:2021dqv}. 
However, the interpretation of parametrized QNM tests of strong-field gravity will require very precise measurements and a more solid theoretical understanding of QNM excitation. 

The paper is structured as follows.  In Sec.~\ref{sec_met} we introduce our theoretical framework and Bayesian data analysis techniques. In Sec.~\ref{sec_app_res} we apply the method to GR and non-GR injections. In Sec.~\ref{sec_dis_con} we discuss our results and outline possible directions for future work.  Throughout the paper we adopt geometrical units ($G=c=1$).

\section{Methodology}\label{sec_met}

In this section we give an overview of the three main ingredients of our method: the parametrized QNM framework based on a Taylor expansion of the perturbation to the GR potential (Sec.~\ref{sec_met_1}), the WKB expansion of the modified potential near the light right (Sec.~\ref{sec_met_WKB}), and the Bayesian inference approach used to address the inverse problem (Sec.~\ref{Bayesian_approach}).

\subsection{Parametrized QNM framework}\label{sec_met_1}

The perturbations of the Schwarzschild black hole were first studied in the odd-parity (axial) case by Regge and Wheeler~\cite{Regge:1957td}, and later extended to the even-parity (polar) case by Zerilli~\cite{Zerilli:1970se}. 
The more general case of Kerr black holes was worked out by Teukolsky~\cite{Teukolsky:1973ha}.

The characteristic frequencies and damping times of the QNMs for rotating and non-rotating black holes in GR have long been known (see~\cite{Kokkotas:1999bd,Nollert:1999ji,Berti:2009kk,Konoplya:2011qq,Pani:2013pma} for reviews). 
More recently, various authors have considered black hole perturbations and QNMs in modified theories of gravity. 
Almost all works are limited to non-rotating or slowly rotating black holes~\cite{Cardoso:2009pk,Molina:2010fb,Blazquez-Salcedo:2016enn,Blazquez-Salcedo:2017txk,Blazquez-Salcedo:2020rhf,Blazquez-Salcedo:2020caw,Franchini:2021bpt}, because finding exact analytical background solutions for arbitrary rotation is not always possible, and separating the perturbation equations is even harder (if at all possible). 
Theory-specific works using the slow-rotation approximation for gravitational perturbations include dynamical Chern-Simons gravity~\cite{Wagle:2021tam,Srivastava:2021imr}, Einstein-dilaton-Gauss-Bonnet gravity~\cite{Pierini:2021jxd,Pierini:2022eim}, and effective-field-theory extensions of GR including higher-derivative terms~\cite{Cano:2020cao,Cano:2021myl}, although there are recent attempts at formulating generalized Teukolsky equations valid for more generic theories and arbitrary rotation~\cite{Li:2022pcy,Hussain:2022ins}.

Since Einstein's theory is very well tested in the weak- and strong-field regimes, from an experimental point of view it is reasonable to treat possible modifications as small deviations from GR. 
Under this assumption, the parametrized framework developed in Refs.~\cite{Cardoso:2019mqo,McManus:2019ulj,Volkel:2022aca} allows for a convenient and efficient calculation of QNMs once the perturbation equations are cast in the form
\begin{equation}\label{eq:mastersystem}
\frac{\dd^2 {\Phi}}{\dd r_*^2} +\left[\om^2 - V(r)\right]{\Phi} = 0 \,, 
\end{equation}
where the radial function $\Phi$ comes from a spherical harmonic decomposition of the perturbed metric or of some other perturbing field, the tortoise coordinate $r_*$ is defined in terms of the areal radius $r$ and the function $f=1-\rh/r$ through $\dd r_* = \dd r/f$, where $\rh$ is the location of the black horizon, and $\om$ is the complex QNM frequency. 
For metric perturbations we can write
\begin{align}
V(r) &=   V_\note{GR} (r) + \sum_{k=0}^{\infty} \alpha^{(k)} \delta V_k(r) \,,
\label{parTaylor}
\end{align}
where $V_\note{GR}$ is either the Regge-Wheeler or Zerilli potential, and
\begin{align}
\delta V_k(r) = \frac{f(r)}{\rh^2}  \left(\frac{\rh}{r} \right)^k\,\label{eq:deltaV}.
\end{align}
If the coefficients $\al^{(k)}$ are small, the QNMs obtained by the solution of Eq.~\eqref{eq:mastersystem} can be approximated, up to second order, by the expression
\begin{equation}\label{eq:omega_mod}
    \om \approx \om^0 + \al^{(k)} d_{(k)} +  
    \frac{1}{2} \alpha^{(k)} \alpha^{(s)} e_{(ks)} \,,
\end{equation}
where $\om^0$ are the GR frequencies, and the coefficients $d_{(k)}$ and $e_{(ks)}$ were first introduced in~\cite{Cardoso:2019mqo,McManus:2019ulj}. 
We compute these coefficients via a continued-fraction method, as in Ref.~\cite{Volkel:2022aca}. 
Note that in general the $\alpha^{(k)}$ can be complex numbers and might depend on the unperturbed QNM frequency, and thus also on the overtone number itself. 
We are also neglecting a quadratic correction term proportional to the possible QNM dependence of the potential correction term, which can be found in Ref.~\cite{McManus:2019ulj}. 
By a redefinition of the field it is possible to reduce the number of parameters in the potential~\cite{Kimura:2020mrh}, but this was only demonstrated in the linear case, and it might not be possible in the quadratic case studied in this work (or in the more general and realistic case of rotating black holes).

The $\alpha^{(k)}$ coefficients must be small enough for the QNM frequencies to be adequately approximated by a quadratic expansion in $\alpha^{(k)}$. 
Therefore we impose the following -- necessary but not sufficient -- ``convergence criterion'' (see~\cite{Cardoso:2019mqo}):
\begin{align}\label{eq:boundcoeff}
\al^{(k)} \ll \al^{(k)}_\note{M} \equiv (k+1) \left(1+\frac{1}{k}\right)^k.
\end{align}

\subsection{Higher-order WKB method}\label{sec_met_WKB}

The QNMs in this work are obtained either numerically (with a continued fraction method) or analytically (through the parametrized QNM framework). 
However, it will be useful to analyze our results by making use of the physical insight derived from a higher-order WKB approximation. 
The WKB approximation is widely used for scattering problems of the form of Eq.~\eqref{eq:mastersystem}, if the potential describes a barrier with a single maximum and suitable asymptotics. 
Within black hole perturbation theory, the leading-order WKB approximation is known as the Schutz-Will formula~\cite{Schutz:1985km}, and it was later generalized to higher orders~\cite{Iyer:1986np,Konoplya:2003ii,Matyjasek:2017psv}. 
The higher-order WKB approximation can be written, schematically, as
\begin{equation}\label{eq:WKBfreq}
    \omega_n^2 = V^{(0)} - \ii \sqrt{-2V^{(2)}}\left(n+\frac{1}{2} \right) + \sum_i{\tilde{\Lambda}_i(n)} \,,
\end{equation}
where $n$ is the overtone number. 
The correction terms in the WKB approximation $\tilde{\Lambda}_i(n)$ are lengthy expressions~\cite{Iyer:1986np,Konoplya:2003ii,Matyjasek:2017psv} (for completeness, we list them in Appendix~\ref{details_WKB} up to third order). 
They involve derivatives $V^{(p)}$ of the effective potential with respect to the tortoise coordinate, evaluated at the maximum of the potential:
\begin{equation}\label{eq:derivV}
V^{(p)} = \frac{\dd^p V}{\dd r_*^p}\Bigg|_{r_\note{MAX}}.
\end{equation}
Increasing the order of the WKB correction introduces even higher-order derivatives of the potential. 
In general the WKB approximation works rather well for the fundamental mode, and it becomes less accurate as $n$ increases~\cite{Konoplya:2019hlu}.

\begin{figure}
\centering
\includegraphics[width=\linewidth]{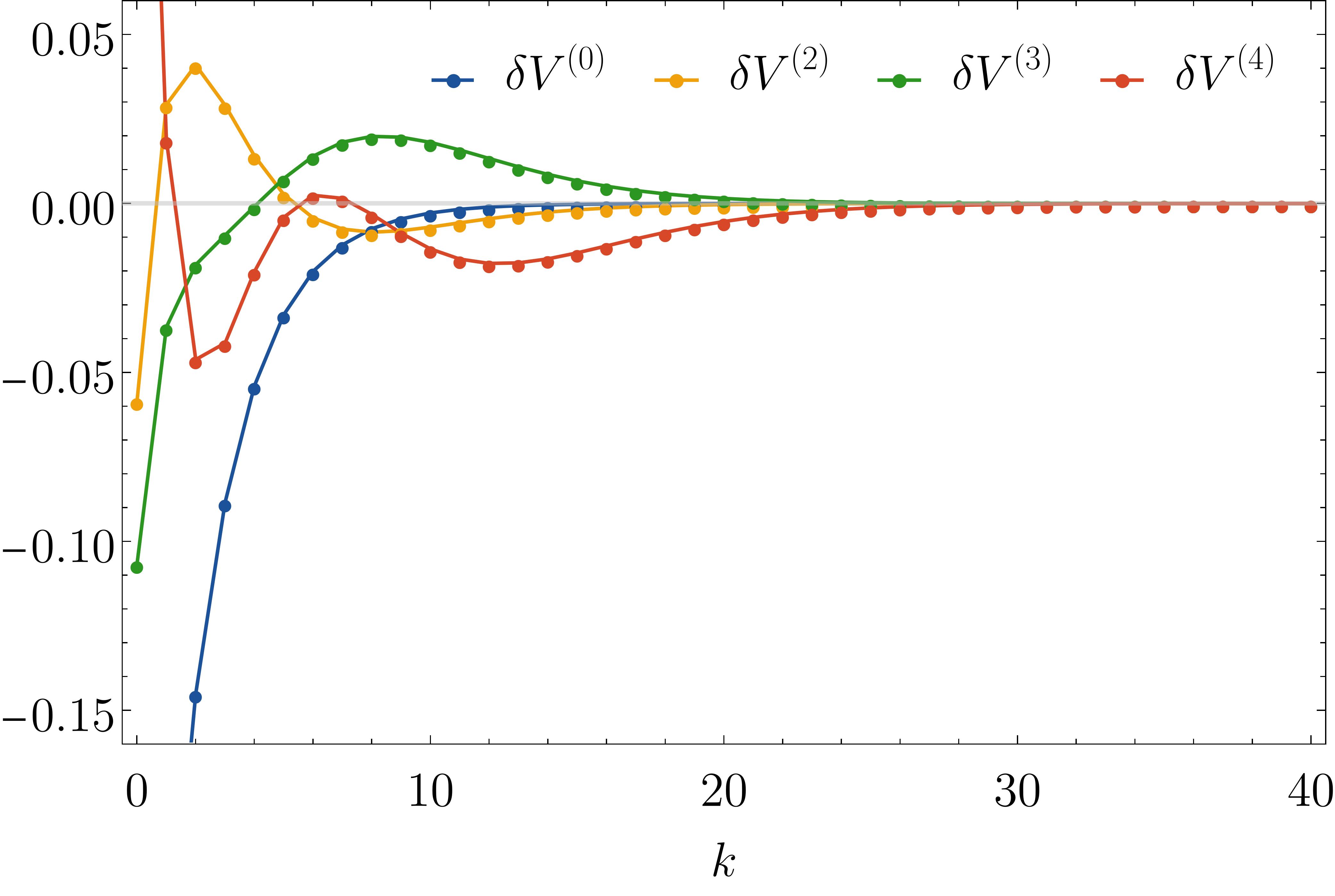}
\caption{Displacement of the derivatives of the effective potential evaluated at the peak for different values of $k$ for axial perturbations with $\ell=2$.  \label{fig:coeff_V}
}
\end{figure}

In the following we revise the calculation of the potential derivatives assuming a perturbative expansion in the coefficients $\al^{(k)}$. 
The maximum of $V$ with respect to the tortoise coordinate is found by solving $\dd V / \dd r_* = 0$, and it is located at $r_\note{MAX} = \rM + \sum_k \al^{(k)}\de r_k$, where $\rM$ is the location of the peak in GR and
\begin{equation}
    \de r_k = - \frac{\de V_k}{V^{(2)}_\note{GR}}\Bigg|_{r=\rM}\,.
\end{equation}
At first order in $\al^{(k)}$, denoting with a $p$ superscript the $p$-th derivative with respect to $r_*$ evaluated at $r_\note{MAX}$, we have
\begin{equation}\label{potential_derivative_expansion}
    V^{(p)} = V^{(p)}_\note{GR} + \sum_{k=0}^{\infty} \al^{(k)} \de V^{(p)}_k.
\end{equation}
In Fig.~\ref{fig:coeff_V} we plot $\de V^{(p)}_k$ for the $\ell=2$ axial potential (up to $p=4$) as a function of $k$ for $k\leq 40$, while in Table~\ref{tab:coeff_V} we list the value of these derivatives for $k\leq 10$.

\begin{table}
\centering
\[
\begin{array}{c|rrrr}
k & \de V^{(0)}_k & \de V^{(2)}_k & \de V^{(3)}_k & \de V^{(4)}_k \\[0.5ex]
\hline
 0 & -0.390 & -0.0585 & -0.107 & 0.256 \\
 1 & -0.238 & 0.0293 & -0.0366 & 0.0189 \\
 2 & -0.145 & 0.0410 & -0.0181 & -0.0462 \\
 3 & -0.0884 & 0.0291 & -0.00937 & -0.0413 \\
 4 & -0.0539 & 0.0141 & -0.000874 & -0.0202 \\
 5 & -0.0329 & 0.00268 & 0.00741 & -0.00407 \\
 6 & -0.0200 & -0.00425 & 0.0140 & 0.00245 \\
 7 & -0.0122 & -0.00756 & 0.0182 & 0.00154 \\
 8 & -0.00745 & -0.00848 & 0.0199 & -0.00319 \\
 9 & -0.00454 & -0.00804 & 0.0196 & -0.00878 \\
 10 & -0.00277 & -0.00697 & 0.0181 & -0.0134 \\
\end{array}
\]
\caption{Derivatives of the effective potential evaluated at the peak for axial perturbations with $\ell=2$ and $k\leq 10$.}
\label{tab:coeff_V}
\end{table}

We can solve for $\om$ and expand the result at first order in the $\al^{(k)}$. 
This yields the following relation for the frequency at third order in the WKB approximation:
\begin{equation}\label{eq:WKBfreq_exp}
\begin{split}
    \om \approx \om_\note{WKB} + \sum_{k=2}^{\infty} \frac{\al^{(k)}}{2\om_\note{WKB}}\bigg[ & \de V^{(0)}_k + \de V^{(2)}_k \frac{\om_\note{WKB}^2 - V^{(0)}_\note{GR}}{2V^{(2)}_\note{GR}} \\
    & +\de\tilde{\La}_2^{k} + \de\tilde{\La}_3^{k}\bigg] \,,
\end{split}
\end{equation}
where $\om_\note{WKB}$ is the frequency evaluated with the WKB method within GR, and $\de\tilde{\La}_i$ are the linear expansions in $\al$ of the coefficients $\tilde{\La}_i$. 
By comparing Eqs.~\eqref{eq:omega_mod} and~\eqref{eq:WKBfreq_exp} order by order, we can map the coefficients $d_{(k)}$ to the derivatives $\de V^{(p)}_k$ of the displaced potential at the peak.

\subsection{Bayesian approach}\label{Bayesian_approach}

Bayesian analysis allows us to relate the observed data $D$ with the underlying parameters $\theta$ of a model, or to compare different models with each other. 
It also allows us to quantitatively include our prior knowledge or assumptions into the analysis, and understand how they affect the posterior through Bayes' theorem
\begin{align}
\mathcal{P}\left(\theta | D \right) = \frac{ \mathcal{P} \left(D | \theta \right) \mathcal{P}\left(\theta \right)}{\mathcal{P}\left( D \right)},
\end{align}
which states that the posterior $\mathcal{P}\left(\theta | D \right)$ is equal to the prior $\mathcal{P}\left(\theta \right)$ times the likelihood $\mathcal{P} \left(D | \theta \right)$, divided by the evidence $\mathcal{P}\left( D \right)$. 
The evidence is often unknown or hard to compute, but it is still possible to compute the posterior via Markov-chain-Monte-Carlo (MCMC) techniques. 
These only require the knowledge of the prior and likelihood, and can be used to directly draw samples from the posterior. 
Standard MCMC techniques become very computationally expensive when the number of sampled parameters is large. 
The parametrized QNM framework of Eq.~\eqref{eq:omega_mod} is very beneficial in this sense, because it speeds up the calculation of the likelihood significantly by avoiding the more involved (and not necessarily always converging) calculations that are necessary in other techniques. 
This results in quick MCMC sampling, allowing us to study a reasonable number of parameters. 

We assume that our likelihood for the unknown parameters $\theta = [\rh, \alpha^{(k)} ]$ is given by
\begin{align}\label{likelihood}
\mathcal{P}(D | \theta) \propto \exp\left[- \frac{1}{2} \vec{h}(\th) \mathbf{C}^{-1} \vec{h}(\th) \right],
\end{align}
where $\mathbf{C}^{-1}$ is the inverse of the covariance matrix $\mathbf{C}$ of the QNM frequencies, and
\begin{equation}
\vec{h}(\th) = \frac{\rh^{0}}{\rh} \vec{\omega}\left(\alpha^{(k)}\right)  - \vec{D}\,,
\end{equation}
where $\rh^0$ is the location of the horizon as it would be inferred in GR. 
We perform this rescaling since we do not know {\it a priori} the correct value for $\rh$. 

The exact form of the correlations depends on the details of the observed binary black hole system, as well as the details of the detector. 
However, by identifying the inverse of the covariance matrix with the Fisher matrix, it is possible to use analytic estimates derived in Ref.~\cite{Berti:2005ys} as an approximation. 
For measurements of the fundamental QNM and of the first overtone ($n=0,1$) the Fisher matrix we need to estimate is a $4 \times 4$ matrix. 
The submatrix that connects the real and imaginary parts of the frequency for a fixed value of $n$ can be approximated with the results of Ref.~\cite{Berti:2005ys}, but the correlations between different $n$'s  (i.e., the $2 \times 2$ off-diagonal blocks) cannot, and therefore we neglect them. 
However, we have analyzed how moderate random correlations between the fundamental mode and the first overtone would change the results, and we did not find substantial qualitative changes. 
We rescale the Fisher matrix so that the real part of the $n=0$ mode is measured with a relative error of $1\,\%$, which yields a relative error of $4.7\,\%$ for its imaginary part. 
Assuming that the $n=1$ overtone is excited with a similar amplitude (an assumption justified by fits to numerical relativity simulations~\cite{Giesler:2019uxc}) yields relative errors of $3.4\,\%$ and $8.2\,\%$ for the real and imaginary part of the overtone frequency, respectively. 
Ongoing work in the literature suggests that the specific model used to extract overtones can have an important impact on interpreting the constraints~\cite{Cheung:2022rbm,Mitman:2022qdl}, 
but this approximate estimate of the Fisher matrix is sufficient for our purposes. 

The location of the black hole horizon $\rh$ affects the QNM spectrum and depends on the underlying theory. 
We assume that $\rh$ is close to its GR value $\rh^0$ and that the uncertainty on $\rh$ is relatively small ($\sigma_{\rh} = 5\,\%$), for example because we have a good estimate of the remnant black hole's mass from the inspiral/merger waveform. 
Then we multiply the likelihood of Eq.~\eqref{likelihood} by an additional factor
\begin{align}
\mathcal{P}_{\rh} \equiv \exp\left[
-\frac{1}{2} \left(\frac{\rh-\rh^0}{\sigma_{\rh}}\right)^2
\right].
\end{align}
This is equivalent to assuming a Gaussian prior for $\rh$ centered around its value $\rh^0$ in GR.

The priors $\mathcal{P}(\al^{(k)} )$ cannot be chosen to be  arbitrary because we are using a parametrized QNM framework. For the perturbative formalism to be valid, we must adopt bounds consistent with Eq.~\eqref{eq:boundcoeff}.
For concreteness, in our analysis we consider two different prior realizations: $\al^{(k)}_\note{P20} \equiv \al^{(k)}_\note{M}/20$ and $\al^{(k)}_\note{P10} \equiv \al^{(k)}_\note{M}/10$, with $k \in [0,10]$. 
Our choice to study 11 parameters for $\al^{(k)}$ guarantees that we can explore a large parameter space to capture modifications to GR, but at the same time we can still perform efficient MCMC sampling. 
To perform the MCMC analysis we use the \textsc{emcee} sampler~\cite{Foreman-Mackey:2012any}, which is based on the affine-invariant ensemble sampler proposed by Goodman and Weare~\cite{2010CAMCS...5...65G}.
Since one would not generically expect that very large $k$ contribute in a dominant way, neglecting higher orders is a reasonable choice.
Throughout this work we assume flat (uniform) priors within these two different bounds. 
Considering different ranges allows us to quantify what aspects of the analysis are sensitive to the choice of priors and which are not. 
By using the continued-fraction numerical code discussed in Ref.~\cite{Volkel:2022aca}, we have checked that the frequencies generated by taking $\al^{(k)} = \pm \al^{(k)}_\note{P20}$, with $k \in [0,10]$, are well approximated by the quadratic expansion. 
However, for certain combinations of random draws from the larger prior range $\al^{(k)} = \pm \al^{(k)}_\note{P10}$ the approximation can become inaccurate. 
Our results below (based on the exact injection) show that this only mildly affects our reconstruction of the perturbative parameters that were used in the parametrized framework for the inference. 

\begin{figure*}
\centering
\includegraphics[width=1.0\linewidth]{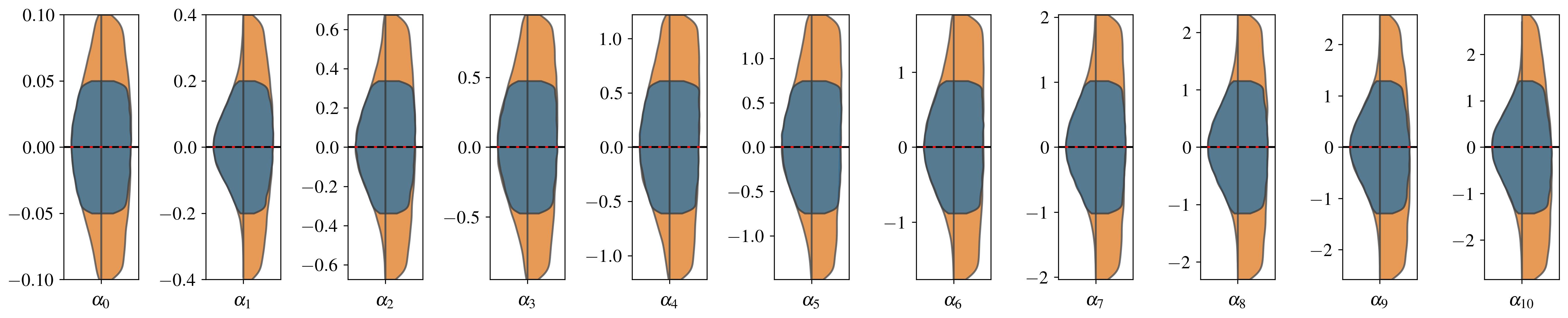}
\caption{MCMC results for simulated observations of the Schwarzschild $n=0,1$ QNMs with relative errors described in the main text, and assuming a Gaussian prior for $r_\mathrm{H}$ corresponding to $5\,\%$ at $1\sigma$ confidence. In each panel, on the left side we plot the posteriors in the optimistic case (only one $\alpha^{(k)}$ is varied at a time), and on the right side we plot the posteriors in the pessimistic case (all $\alpha^{(k)}$'s are varied simultaneously). The blue (orange) color corresponds to a prior range of $\al^{(k)}_\note{P20}$ ($\al^{(k)}_\note{P10}$).}\label{alpha_GR}
\end{figure*}

\begin{figure*}
\centering
\includegraphics[width=1.0\linewidth]{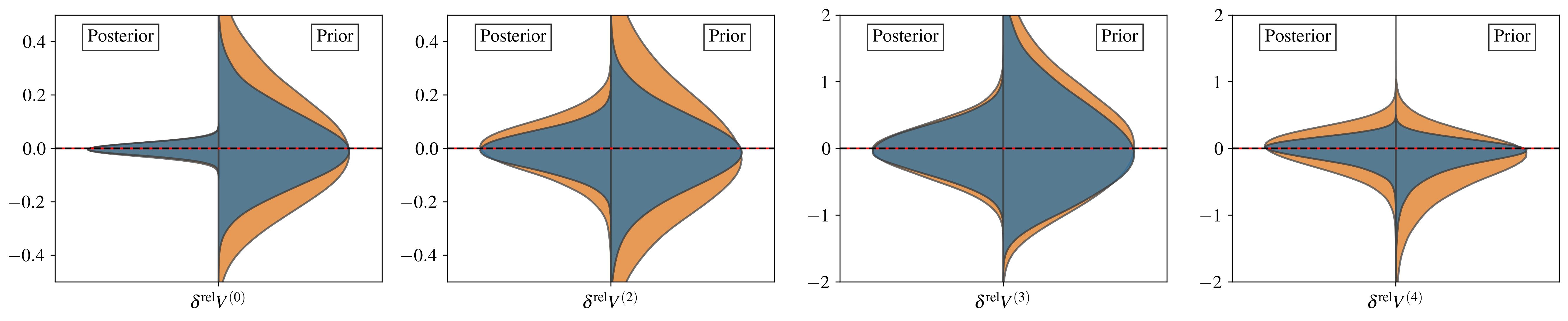}
\caption{Same MCMC results as in Fig.~\ref{alpha_GR}, but here we sample the relative errors of various derivatives of the effective potential $\delta^\mathrm{rel}V^{(n)}$ with respect to the tortoise coordinate evaluated at the maximum. The right side of each violin plot shows the sampling from the prior distribution of all $\alpha^{(k)} $ simultaneously, with colors label the different $\alpha^{(k)}$ prior ranges: $\al^{(k)}_\note{P20}$ (blue) and $\al^{(k)}_\note{P10}$ (orange). 
The left side of each violin plot shows the sampling from the posterior distributions, with colors labeling the prior assumptions. 
We mark the injections (here corresponding to a Schwarzschild black hole) by red horizontal lines.}
\label{wkb_GR}
\end{figure*}

\section{Results}\label{sec_app_res}

Using the techniques of the previous section, we now study two complementary settings.  
In the first setting (Sec.~\ref{res_GR}) we perform injections assuming that GR is the correct theory of gravity. 
In the second setting (Sec.~\ref{res_non-GR}) we assume a hypothetical modification with two non-zero deviation parameters -- a nontrivial, but tractable case. 

Within each setting we use the (complex) $l=2,$ $n=0,1$ QNM frequencies as hypothetical observations, with errors estimated by the Fisher matrix formalism as described in Sec.~\ref{Bayesian_approach}. 
To explore the impact of different priors on the posteriors, we always use two different prior ranges ($\al^{(k)}_\note{P20}$ and $\al^{(k)}_\note{P10}$). 

Within each of these two settings, we will address the inverse problem in two steps. We will first compare the posteriors for the parameters $\alpha^{(k)}$ in the optimistic scenario (where only one parameter varies) against the pessimistic scenario (where all parameters are varied simultaneously). 
Then, from the reconstruction of the potential, we will evaluate $V^{(p)}$ and compare it with the corresponding value $V^{(p)}_\note{GR}$ in GR by computing the relative difference
\begin{align}\label{potential_rel}
\delta^\mathrm{rel} V^{(p)} = \frac{V^{(p)}(\rh, \alpha^{(k)})-V^{(p)}_\note{GR}}{V^{(p)}_\note{GR}}.
\end{align}

\subsection{Results for a GR injection}\label{res_GR}

First we present our results under the assumption that GR is the correct theory of gravity. 

The violin plots in Fig.~\ref{alpha_GR} show the optimistic (left side) vs pessimistic (right side) posteriors for the Taylor expansion coefficients $\alpha^{(k)}$in Eq.~\eqref{parTaylor}. 
Different colors correspond to posteriors obtained with the two different (flat) prior ranges. 
In the optimistic case, it is possible to constrain all parameters independently of the chosen prior range. 
While the quantitative details of each optimistic bound are different and their widths increase with $k$ (as expected), they are qualitatively the same for all $k$. 
The situation is clearly different in the pessimistic case. 
Here the posterior distributions of all $\alpha^{(k)}$ have support at the prior boundaries, and they become broader when the prior range is increased. 
Since we allow for more parameters (12 in total: 11 $\alpha^{(k)}$'s and $\rh$) than observed QNMs (4, {\it i.e.}, 2 complex frequencies), this is not surprising. 

In Fig.~\ref{wkb_GR} we show the results for the WKB deviation coefficients $\delta^\mathrm{rel} V^{(n)}$ defined in Eq.~\eqref{potential_rel}. 
Since flat priors for $\alpha^{(k)}$ do not correspond to flat priors for $\delta^\mathrm{rel} V^{(n)} $, it is important to compare the posteriors with the priors. 
We want to understand if QNM frequencies contain information on the $\delta^\mathrm{rel} V^{(n)}$ coefficients, or if the inference is prior-dominated (in which case the QNMs would not be informative). 
We focus on the pessimistic posteriors, since the more general (pessimistic) assumption allows us to draw conservative conclusions. 
As usual, we show the two different prior ranges using different colors. 

Quite remarkably, the posteriors for $\delta^\mathrm{rel} V^{(0)}$ are very robust, and they do not depend on the prior range choice. 
The posteriors for $\delta^\mathrm{rel} V^{(2)}$ and $\delta^\mathrm{rel} V^{(3)}$ are still more informative than the priors, but the QNM frequencies do not provide as much information as they do in the case of $\delta^\mathrm{rel} V^{(0)}$, and the measurement is even less informative in the case of $\delta^\mathrm{rel} V^{(4)}$. 
This is one of the main results of this paper: the ``change of basis'' from the $\alpha^{(k)}$'s to the $\delta^\mathrm{rel} V^{(n)}$'s (variables related to the light ring) is very effective, because QNMs carry physical information about the potential and its derivatives at the peak. 
This is in agreement with the conclusions of Ref.~\cite{Volkel:2022aca}.

Note that the derivation of these results is fully independent of the WKB approximation: we never use the WKB approximation to compute the QNMs, but rather we use Leaver's method to compute the injected QNM frequencies, and  the parametrized framework of Eq.~\eqref{eq:omega_mod} to model their deviations from GR. 
The full MCMC analysis can capture correlations between the deviation parameters, as long as they are small enough to ensure the validity of the parametrized framework.
 
\begin{figure*}
\centering
\includegraphics[width=1.0\linewidth]{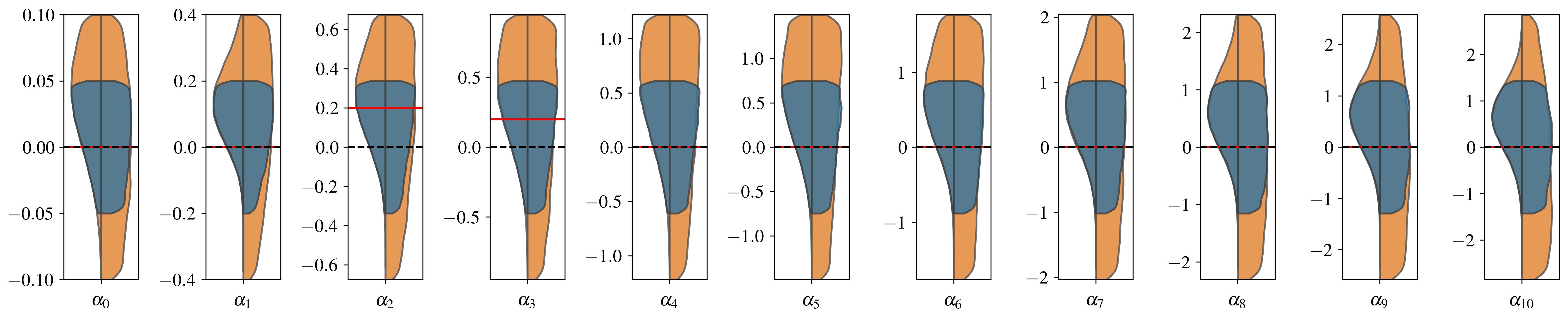}
\caption{MCMC results for simulated observations of a modified Schwarzschild $n=0,1$ QNMs with relative errors described in the main text and assuming a prior for $r_\mathrm{H}$ corresponding to $5\,\%$ at $1\sigma$ confidence. The injected values (shown as horizontal red lines) are $\alpha^{(2)} = \alpha^{(3)} = 0.2$, with all other values of $\alpha^{(k)}$ set equal to zero. From left to right we show the potential coefficients $\alpha^{(k)}$. In each panel, on the left side we plot the posteriors in the optimistic case (only one $\alpha^{(k)}$ is varied at a time), and on the right side we plot the posteriors in the pessimistic case (all $\alpha^{(k)}$'s are varied simultaneously). The blue (orange) color corresponds to a prior range of $\al^{(k)}_\note{P20}$ ($\al^{(k)}_\note{P10}$).} \label{alpha_a02_k25}
\end{figure*}

\begin{figure*}
\centering
\includegraphics[width=1.0\linewidth]{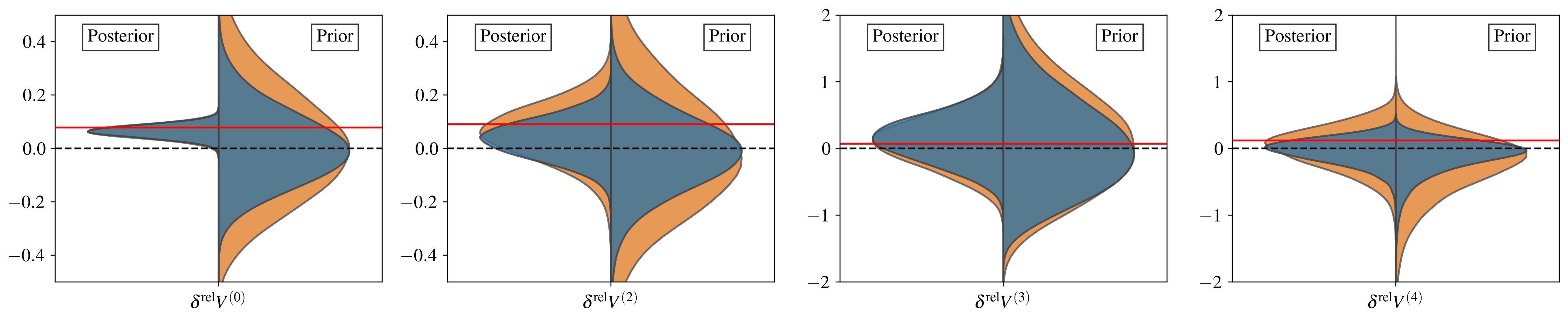}
\caption{Same MCMC results as in Fig.~\ref{alpha_a02_k25}, but here we sample the relative errors of various derivatives of the effective potential $\delta^\mathrm{rel}V^{(n)}$ with respect to the tortoise coordinate evaluated at the maximum. The right side of each violin plot shows the sampling from the prior distribution of all $\alpha^{(k)} $ simultaneously. The colors label the different $\alpha^{(k)}$ prior ranges, $\al^{(k)}_\note{P20}$ (blue) and $\al^{(k)}_\note{P10}$ (orange).
The left side of each violin plot shows the sampling from the posterior distributions, with colors labeling the prior assumptions. 
We mark the injections by horizontal red lines.
}
\label{wkb_a02_k23}
\end{figure*}

\subsection{Results for a non-GR injection}\label{res_non-GR}

As an example of a possible non-GR injection, we consider the hypothetical case in which $\alpha^{(2)} = \alpha^{(3)} = 0.2$, while all other $\alpha^{(k)}$'s are set to zero. 
We compute the QNM frequency with Leaver's method, while we use the parametrized framework for the MCMC analysis. 

In Fig.~\ref{alpha_a02_k25} we report the MCMC results, following the same notation and conventions as in Fig.~\ref{alpha_GR}. 
The injected values of $\alpha^{(k)}$ are marked by horizontal red lines. 
The plot shows that it is not possible to infer the correct values for any $\alpha^{(k)}$, except perhaps for $k=2$ and (less clearly) for $k=3$. The recovered values for all other $\alpha^{(k)}$'s peak away from their injected value. 
Some of the small-$k$ posteriors have strong support at the smaller prior limit, but they are well captured by the larger prior limit. 

Given that these results are qualitatively very different from the GR injections, it is quite remarkable that the WKB-motivated constraints on the potential near the peak are as robust as before. 
In Fig.~\ref{wkb_a02_k23} we show that it is indeed possible to recover the non-GR injections (shown as red horizontal lines). 
The posterior of the dominant term $\delta^\mathrm{rel} V^{(0)}$ has tight support away from GR, while $\delta^\mathrm{rel} V^{(2)}$ is broader and has some overlap with the GR hypothesis. 
The higher derivatives for this injection are very close to their GR value, and the corresponding violin plots are almost indistinguishable from those of Fig.~\ref{wkb_GR}.

\subsection{Theory-specific approach: a simple example}\label{res_3}

A theory-specific analysis can, in principle, be carried out in different ways. 
We could avoid making use of the parametrized framework and base it on a full MCMC analysis involving all free parameters of the theory. 
We could also, in principle, use the parametrized framework with the theory-predicted values of $\alpha^{(k)}$, which could depend on one (or multiple) coupling constants of the modified theory of gravity of interest. 
Instead of repeating a full MCMC analysis, here we show that the posteriors for $\delta^\mathrm{rel} V^{(0)}$ in the non-GR example above are already very informative when used in a ``post-processing'' analysis. 

Let us assume,  for simplicity, that the hypothetical modified theory predicts two non-zero deviation parameters, related to the only unknown parameter of the theory $\ze$ as follows: 
\begin{align}\label{coupling_function}
\alpha^{(2)}(\ze)  = \alpha^{(3)} (\ze) = \ze.
\end{align}
We use this trivial example just for illustrative purposes, but no fundamental limitation prevents us from relaxing these assumptions to allow for more than two non-zero deviation parameters, or to allow for a more complicated dependence of these parameters on $\ze$. 
In fact, we would find qualitatively similar results if we assumed that $\alpha^{(3)}$ is suppressed by a small factor $\epsilon$. 

\begin{figure}
\centering
\includegraphics[width=\linewidth]{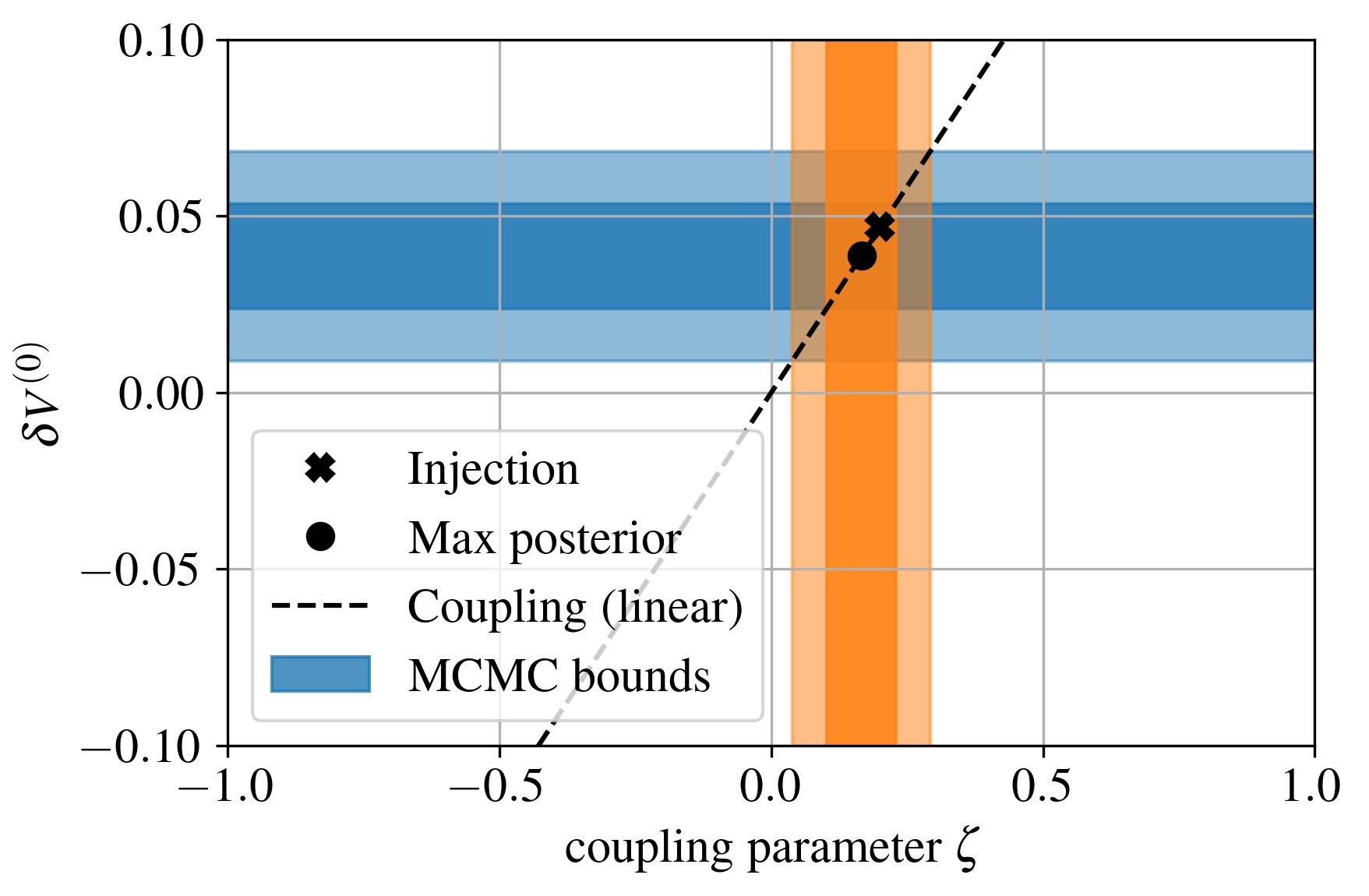}
\caption{Post-processing analysis using the posterior of $V^{(0)}$. Blue areas show the $1\sigma$ and $2\sigma$ confidence levels of the theory-agnostic analysis, while the orange areas label the corresponding bounds on the theory-specific parameter $\zeta$. The black dashed line indicates the linear scaling predicted by Eq.~\eqref{potential_derivative_expansion}. The black dot shows the maximum of the posterior, and the cross is the injected value. \label{theory_specific}
}
\end{figure}

By using Eq.~\eqref{potential_derivative_expansion} we can find the linear approximation for the modified potential $V^{(0)}(\ze)$. 
The GR potential $V^{(0)}$ is known, so we can express the posteriors $\delta^\mathrm{rel} V^{(0)}$ in Fig.~\ref{wkb_a02_k23} (left side of the first violin plot) in terms of $V^{(0)}$. 
The posteriors are well approximated by a Gaussian, whose mean $\mu_{V}$ and width $\sigma_V$ we can fit numerically. 
Then we can insert the value of $\mu_{V}$ fitted  to $V^{(0)}$ into Eq.~\eqref{potential_derivative_expansion}, and invert it to find the most likely value of $\ze$, which we will denote by $\mu_\ze$. 
We can also estimate the $1\sigma$ and $2\sigma$ errors by repeating the inversion for $\mu_V \pm \sigma_V$ and $\mu_V \pm 2\sigma_V$. 
In Fig.~\ref{theory_specific} we show the results of this procedure. 
The injected value of the coupling parameter $\zeta$ is well within the $1\sigma$ constraint. 
The values of $\mu_\ze$ and $\sigma_\ze$ can also be used to define the theory-specific bounds on $\alpha^{(2)}(\ze)$ and $\alpha^{(3)}(\ze)$ in a similar way. 

\begin{figure}
\centering
\includegraphics[width=\linewidth]{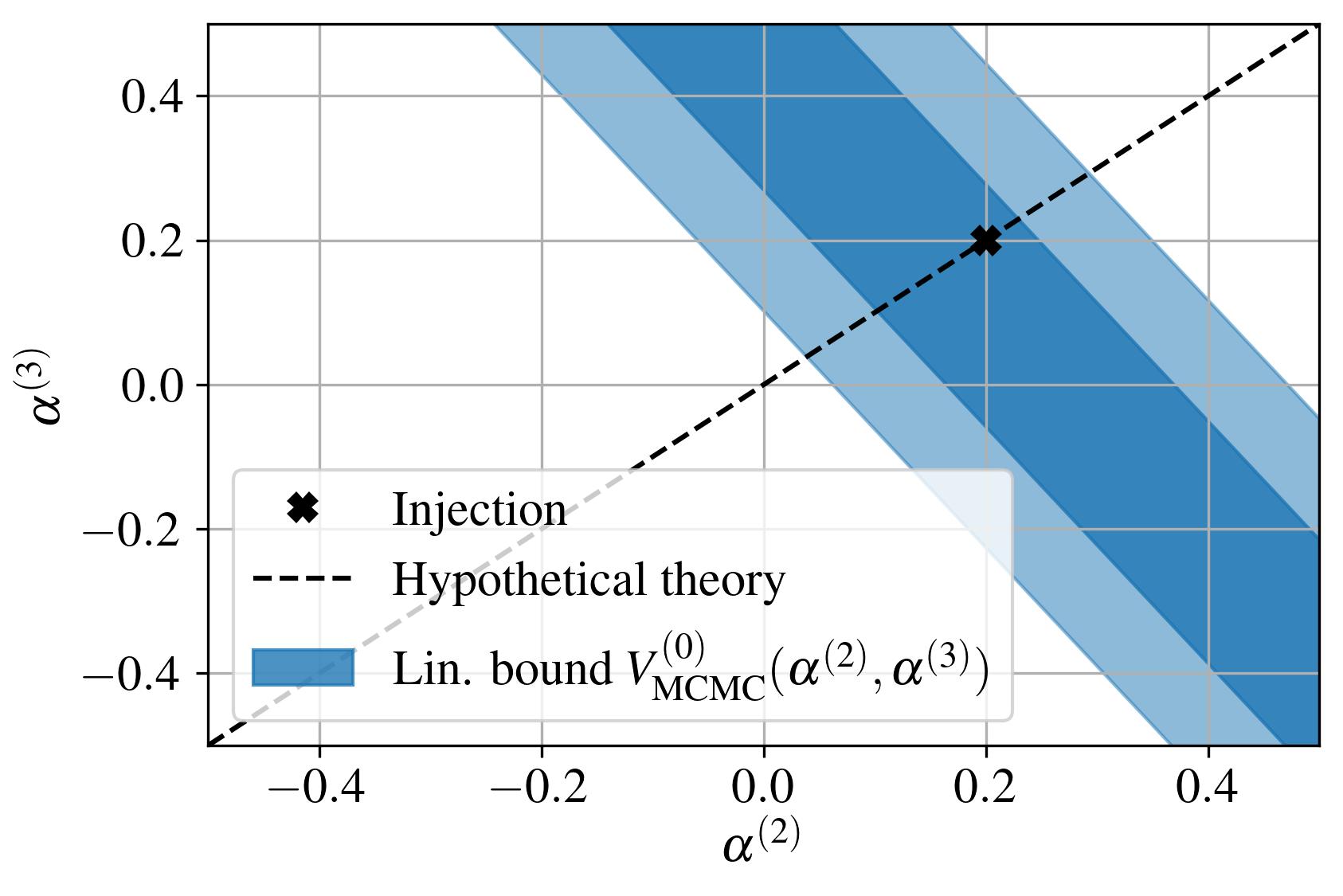}
\caption{Post-processing analysis using the posterior of $V^{(0)}$. Blue areas show the $1\sigma$ and $2\sigma$ confidence levels on the functional relation $\alpha^{(3)}(\alpha^{(2)}, V^{(0)})$. 
The black dashed line indicates the linear relation assumed in the hypothetical theory. 
The black dot is the intersection of the maximum of the posterior with the theory-specific relation, while the cross is the injected value. \label{theory_specific_2}
}
\end{figure}

If we do not know the precise form of $\alpha^{(2)}(\ze), \alpha^{(3)}(\ze)$ ({\it i.e.}, in a theory-agnostic approach), we could treat them as independent variables in a post-processing analysis. 
From Eq.~\eqref{potential_derivative_expansion} we can find $\alpha^{(3)}$ as a function of $\alpha^{(2)}$ and $V^{(0)}$, with the results shown in Fig.~\ref{theory_specific_2}. 
Once again, the injected value is within the $1\sigma$ confidence levels. 

Finally, we could repeat the analysis by using the posteriors for $V^{(2)}$, which would in principle break the two-parameter degeneracy of the agnostic case. 
In practice, we find that the additional constraint is almost parallel to the constraint from $V^{(0)}$ and that it only excludes large values of the deviation parameters, for which the approximations do not hold anyway. 
These conclusions depend on the specific combination $\alpha^{(2)}, \alpha^{(3)}$ that we have chosen in our example, and they may be qualitatively different for other combinations of the deviation parameters. 

\section{Discussion and conclusions}\label{sec_dis_con}

In general, one may hope that modifications to the perturbation potential proportional to $\alpha^{(k)} \left(  \rh / r \right)^k$ would yield smaller corrections to the QNM spectrum as $k$ grows, as long as the coefficients $\alpha^{(k)}$ are comparable in order of magnitude. 
In fact, while this trend is present, the contributions from terms with $k \geq 2$ decay quite slowly. 
This lack of a strong hierarchy implies that, in general, it will be very difficult to recover the individual coefficients $\alpha^{(k)}$, at least in the absence of an underlying theory-specific ansatz. 
This was, indeed, the conclusion of previous work on parametrized ringdown using a principal component analysis~\cite{Volkel:2022aca}.

Here we confirm this conclusion by exploring two specific cases: a GR QNM injection, and a non-GR injection.
In both cases we look at two extreme scenarios: in the optimistic scenario we vary a single coefficient $\alpha^{(k)}$ at a time, and in the pessimistic scenario we let multiple $\alpha^{(k)}$'s (with $k=0,\dots10$) vary simultaneously. 
In both scenarios we find that it is difficult to constrain the individual $\alpha^{(k)}$'s, as expected. 
We study a non-GR injection in which we assume that only two of the $\alpha^{(k)}$'s are non-zero. 
In the optimistic scenario we find that the posteriors of all of the $\alpha^{(k)}$'s disfavor GR, but only one of them is close to the correct injected value. 
Therefore it is difficult to identify which non-zero $\alpha^{(k)}$'s are present in the data without having more precise data, or additional (theory-specific) criteria to interpret the posteriors.

The main conclusion of this work is that the problem can be by-passed by exploiting the well-known relation between QNM frequencies and the properties of the perturbation potential at the light ring. 
This relation suggests that the complex correlations present when many $\alpha^{(k)}$'s are varied simultaneously can be understood by relating the $\alpha^{(k)}$'s to the value of the potential and its derivatives at the light ring. 
It is possible to obtain very robust constraints on the potential and its derivatives at the light ring, even in the pessimistic case in which many $\alpha^{(k)}$'s are varied simultaneously. 
More precisely, a measurement of the fundamental mode and of the first overtone can constrain quite precisely the value of the potential and of its second derivative evaluated at the peak. 
As demonstrated, this conclusion is not limited to GR injections, but it is also valid for non-GR injections. 

We analyze, as a proof of principle, a hypothetical theory characterized by a single parameter $\zeta$, such that all non-zero deviations have the form $\alpha^{(k)}(\zeta)$. 
Through a simple ``post-processing'' analysis based on the inferred posterior for the value of the potential at the maximum, and assuming that only two of the $\alpha^{(k)}$'s are non-zero, we can recover the injected parameter $\zeta$ and estimate confidence intervals for each of the $\alpha^{(k)}$'s. 
This demonstrates that MCMC constraints on the value of the potential at the maximum can be used to find theory-specific properties, without rerunning a full MCMC analysis.

The intimate link between QNM frequencies and the potential at the light ring implies that there is a close relation between strong-gravity tests using black hole spectroscopy and the black hole shadow observations by the Event Horizon Telescope collaboration~\cite{EventHorizonTelescope:2019dse,EventHorizonTelescope:2019ggy,EventHorizonTelescope:2021dqv}. 
In analogy with QNMs, the strong-field character of the shadow (which depends on the geometry near the light ring) also implies that the individual parameters in a PN expansion of the metric are hard to constrain, especially when many of them are allowed to vary simultaneously~\cite{Volkel:2020xlc}. 
Black hole shadow measurements can still constrain linear combinations of the metric and its derivative at the light ring~\cite{Volkel:2020xlc}, in a way which is closely reminiscent of the WKB results reported in this work.

In closing, it is important to discuss some caveats and possible future extensions of this work. 
The parametrized framework~\cite{Cardoso:2019mqo,McManus:2019ulj} is only valid for non-rotating or slowly rotating black holes, but current gravitational-wave observations involve rotating black holes. 
More theory-dependent studies of QNMs for rotating black holes in modified theories of gravity are necessary to overcome this nontrivial obstacle, and to understand how a generalized parametrized framework can effectively be implemented and constrained. 
The increase in complexity resulting from the (generally nonseparable) perturbation equations in modified gravity implies that theory-specific tests may be particularly valuable, because they typically involve a small number of free parameters.

\acknowledgments
N.~Franchini would like to thank the Johns Hopkins University for hospitality during the early stages of this work.
S.~H.~V\"olkel, N.~Franchini and E.~Barausse acknowledge financial support provided under the European Union's H2020 ERC Consolidator Grant ``GRavity from Astrophysical to Microscopic Scales'' grant agreement no. GRAMS-815673. 
This work was supported by the EU Horizon 2020 Research and Innovation Programme under the Marie Sklodowska-Curie Grant Agreement No. 101007855. 
S.~H.~V\"olkel acknowledges funding from the Deutsche Forschungsgemeinschaft (DFG) - project number: 386119226. 
E.~Berti is supported by NSF Grants No. AST-2006538, PHY-2207502, PHY-090003 and PHY20043, and NASA Grants No. 19-ATP19-0051, 20-LPS20- 0011 and 21-ATP21-0010. 
This research project was conducted using computational resources at the Maryland Advanced Research Computing Center (MARCC). 
The authors also acknowledge the Texas Advanced Computing Center (TACC) at The University of Texas at Austin for providing HPC resources that have contributed to the research results reported within this paper. URL: \url{http://www.tacc.utexas.edu}~\citep{10.1145/3311790.3396656}.

\appendix

\section{Details for higher order WKB method}\label{details_WKB}

In this appendix we report some of the lengthy WKB expressions used in the main text. 
For convenience, we report again the WKB formula~\eqref{eq:WKBfreq}, up to third order:
\begin{equation}
    \omega_n^2 \simeq V^{(0)} - \ii \sqrt{-2V^{(2)}}\al_n + \tilde{\Lambda}_2(n) + \tilde{\Lambda}_3(n)
\end{equation}
Here $n$ is the QNM overtone number, $\al_n = n+1/2$, and the higher-order corrections read
\begin{align}
    \tilde{\La}_2 = & \frac{1}{8} \left[ \left( \frac{1}{4} + \al_n^2 \right) \frac{V^{(4)}}{V^{(2)}} 
    - \frac{1}{36} \left( 7 + 60 \al_n^2 \right) \frac{{V^{(3)}}^2}{{V^{(2)}}^2} \right]\,, \\
    \tilde{\La}_3 = & -\frac{\ii \al_n}{96\sqrt{-2V^{(2)}}} \Bigg[  \frac{5}{72}\left(77+188\al_n^2\right)\frac{{V^{(3)}}^4}{{V^{(2)}}^4}  \notag \\
    & -\frac{1}{4}\left(51+100\al_n^2\right)\frac{{V^{(3)}}^2V^{(4)}}{{V^{(2)}}^3} \notag \\
    & + \frac{1}{24} \left(67 + 68 \al_n^2 \right) \frac{{V^{(4)}}^2}{{V^{(2)}}^2} \notag \\ 
    & + \frac{1}{3} \left( 19+28\al_n^2 \right) \frac{{V^{(3)}}V^{(5)}}{{V^{(2)}}^2} \notag \\
    & - \frac{1}{3} \left(5+4\al_n^2 \right) \frac{{V^{(6)}}}{{V^{(2)}}} \Bigg]\,.
\end{align}

\bibliography{literature}

\begin{thebibliography}{96}%
\makeatletter
\providecommand \@ifxundefined [1]{%
 \@ifx{#1\undefined}
}%
\providecommand \@ifnum [1]{%
 \ifnum #1\expandafter \@firstoftwo
 \else \expandafter \@secondoftwo
 \fi
}%
\providecommand \@ifx [1]{%
 \ifx #1\expandafter \@firstoftwo
 \else \expandafter \@secondoftwo
 \fi
}%
\providecommand \natexlab [1]{#1}%
\providecommand \enquote  [1]{``#1''}%
\providecommand \bibnamefont  [1]{#1}%
\providecommand \bibfnamefont [1]{#1}%
\providecommand \citenamefont [1]{#1}%
\providecommand \href@noop [0]{\@secondoftwo}%
\providecommand \href [0]{\begingroup \@sanitize@url \@href}%
\providecommand \@href[1]{\@@startlink{#1}\@@href}%
\providecommand \@@href[1]{\endgroup#1\@@endlink}%
\providecommand \@sanitize@url [0]{\catcode `\\12\catcode `\$12\catcode
  `\&12\catcode `\#12\catcode `\^12\catcode `\_12\catcode `\%12\relax}%
\providecommand \@@startlink[1]{}%
\providecommand \@@endlink[0]{}%
\providecommand \url  [0]{\begingroup\@sanitize@url \@url }%
\providecommand \@url [1]{\endgroup\@href {#1}{\urlprefix }}%
\providecommand \urlprefix  [0]{URL }%
\providecommand \Eprint [0]{\href }%
\providecommand \doibase [0]{http://dx.doi.org/}%
\providecommand \selectlanguage [0]{\@gobble}%
\providecommand \bibinfo  [0]{\@secondoftwo}%
\providecommand \bibfield  [0]{\@secondoftwo}%
\providecommand \translation [1]{[#1]}%
\providecommand \BibitemOpen [0]{}%
\providecommand \bibitemStop [0]{}%
\providecommand \bibitemNoStop [0]{.\EOS\space}%
\providecommand \EOS [0]{\spacefactor3000\relax}%
\providecommand \BibitemShut  [1]{\csname bibitem#1\endcsname}%
\let\auto@bib@innerbib\@empty
\bibitem [{\citenamefont {Abbott}\ \emph
  {et~al.}(2016{\natexlab{a}})\citenamefont {Abbott} \emph
  {et~al.}}]{LIGOScientific:2016aoc}%
  \BibitemOpen
  \bibfield  {author} {\bibinfo {author} {\bibfnamefont {B.~P.}\ \bibnamefont
  {Abbott}} \emph {et~al.} (\bibinfo {collaboration} {LIGO Scientific,
  Virgo}),\ }\bibfield  {title} {\enquote {\bibinfo {title} {{Observation of
  Gravitational Waves from a Binary Black Hole Merger}},}\ }\href {\doibase
  10.1103/PhysRevLett.116.061102} {\bibfield  {journal} {\bibinfo  {journal}
  {Phys. Rev. Lett.}\ }\textbf {\bibinfo {volume} {116}},\ \bibinfo {pages}
  {061102} (\bibinfo {year} {2016}{\natexlab{a}})},\ \Eprint
  {http://arxiv.org/abs/1602.03837} {arXiv:1602.03837 [gr-qc]} \BibitemShut
  {NoStop}%
\bibitem [{\citenamefont {Abbott}\ \emph
  {et~al.}(2019{\natexlab{a}})\citenamefont {Abbott} \emph
  {et~al.}}]{LIGOScientific:2018mvr}%
  \BibitemOpen
  \bibfield  {author} {\bibinfo {author} {\bibfnamefont {B.~P.}\ \bibnamefont
  {Abbott}} \emph {et~al.} (\bibinfo {collaboration} {LIGO Scientific,
  Virgo}),\ }\bibfield  {title} {\enquote {\bibinfo {title} {{GWTC-1: A
  Gravitational-Wave Transient Catalog of Compact Binary Mergers Observed by
  LIGO and Virgo during the First and Second Observing Runs}},}\ }\href
  {\doibase 10.1103/PhysRevX.9.031040} {\bibfield  {journal} {\bibinfo
  {journal} {Phys. Rev. X}\ }\textbf {\bibinfo {volume} {9}},\ \bibinfo {pages}
  {031040} (\bibinfo {year} {2019}{\natexlab{a}})},\ \Eprint
  {http://arxiv.org/abs/1811.12907} {arXiv:1811.12907 [astro-ph.HE]}
  \BibitemShut {NoStop}%
\bibitem [{\citenamefont {Abbott}\ \emph
  {et~al.}(2021{\natexlab{a}})\citenamefont {Abbott} \emph
  {et~al.}}]{LIGOScientific:2020ibl}%
  \BibitemOpen
  \bibfield  {author} {\bibinfo {author} {\bibfnamefont {R.}~\bibnamefont
  {Abbott}} \emph {et~al.} (\bibinfo {collaboration} {LIGO Scientific,
  Virgo}),\ }\bibfield  {title} {\enquote {\bibinfo {title} {{GWTC-2: Compact
  Binary Coalescences Observed by LIGO and Virgo During the First Half of the
  Third Observing Run}},}\ }\href {\doibase 10.1103/PhysRevX.11.021053}
  {\bibfield  {journal} {\bibinfo  {journal} {Phys. Rev. X}\ }\textbf {\bibinfo
  {volume} {11}},\ \bibinfo {pages} {021053} (\bibinfo {year}
  {2021}{\natexlab{a}})},\ \Eprint {http://arxiv.org/abs/2010.14527}
  {arXiv:2010.14527 [gr-qc]} \BibitemShut {NoStop}%
\bibitem [{\citenamefont {Abbott}\ \emph
  {et~al.}(2021{\natexlab{b}})\citenamefont {Abbott} \emph
  {et~al.}}]{LIGOScientific:2021djp}%
  \BibitemOpen
  \bibfield  {author} {\bibinfo {author} {\bibfnamefont {R.}~\bibnamefont
  {Abbott}} \emph {et~al.} (\bibinfo {collaboration} {LIGO Scientific, VIRGO,
  KAGRA}),\ }\bibfield  {title} {\enquote {\bibinfo {title} {{GWTC-3: Compact
  Binary Coalescences Observed by LIGO and Virgo During the Second Part of the
  Third Observing Run}},}\ }\href@noop {} {\  (\bibinfo {year}
  {2021}{\natexlab{b}})},\ \Eprint {http://arxiv.org/abs/2111.03606}
  {arXiv:2111.03606 [gr-qc]} \BibitemShut {NoStop}%
\bibitem [{\citenamefont {Nitz}\ \emph {et~al.}(2021)\citenamefont {Nitz},
  \citenamefont {Kumar}, \citenamefont {Wang}, \citenamefont {Kastha},
  \citenamefont {Wu}, \citenamefont {Sch\"afer}, \citenamefont {Dhurkunde},\
  and\ \citenamefont {Capano}}]{Nitz:2021zwj}%
  \BibitemOpen
  \bibfield  {author} {\bibinfo {author} {\bibfnamefont {Alexander~H.}\
  \bibnamefont {Nitz}}, \bibinfo {author} {\bibfnamefont {Sumit}\ \bibnamefont
  {Kumar}}, \bibinfo {author} {\bibfnamefont {Yi-Fan}\ \bibnamefont {Wang}},
  \bibinfo {author} {\bibfnamefont {Shilpa}\ \bibnamefont {Kastha}}, \bibinfo
  {author} {\bibfnamefont {Shichao}\ \bibnamefont {Wu}}, \bibinfo {author}
  {\bibfnamefont {Marlin}\ \bibnamefont {Sch\"afer}}, \bibinfo {author}
  {\bibfnamefont {Rahul}\ \bibnamefont {Dhurkunde}}, \ and\ \bibinfo {author}
  {\bibfnamefont {Collin~D.}\ \bibnamefont {Capano}},\ }\bibfield  {title}
  {\enquote {\bibinfo {title} {{4-OGC: Catalog of gravitational waves from
  compact-binary mergers}},}\ }\href@noop {} {\  (\bibinfo {year} {2021})},\
  \Eprint {http://arxiv.org/abs/2112.06878} {arXiv:2112.06878 [astro-ph.HE]}
  \BibitemShut {NoStop}%
\bibitem [{\citenamefont {Olsen}\ \emph {et~al.}(2022)\citenamefont {Olsen},
  \citenamefont {Venumadhav}, \citenamefont {Mushkin}, \citenamefont {Roulet},
  \citenamefont {Zackay},\ and\ \citenamefont {Zaldarriaga}}]{Olsen:2022pin}%
  \BibitemOpen
  \bibfield  {author} {\bibinfo {author} {\bibfnamefont {Seth}\ \bibnamefont
  {Olsen}}, \bibinfo {author} {\bibfnamefont {Tejaswi}\ \bibnamefont
  {Venumadhav}}, \bibinfo {author} {\bibfnamefont {Jonathan}\ \bibnamefont
  {Mushkin}}, \bibinfo {author} {\bibfnamefont {Javier}\ \bibnamefont
  {Roulet}}, \bibinfo {author} {\bibfnamefont {Barak}\ \bibnamefont {Zackay}},
  \ and\ \bibinfo {author} {\bibfnamefont {Matias}\ \bibnamefont {Zaldarriaga}}
  (\bibinfo {collaboration} {LIGO Scientific Collaboration, the Virgo}),\
  }\bibfield  {title} {\enquote {\bibinfo {title} {{New binary black hole
  mergers in the LIGO-Virgo O3a data}},}\ }\href {\doibase
  10.1103/PhysRevD.106.043009} {\bibfield  {journal} {\bibinfo  {journal}
  {Phys. Rev. D}\ }\textbf {\bibinfo {volume} {106}},\ \bibinfo {pages}
  {043009} (\bibinfo {year} {2022})},\ \Eprint
  {http://arxiv.org/abs/2201.02252} {arXiv:2201.02252 [astro-ph.HE]}
  \BibitemShut {NoStop}%
\bibitem [{\citenamefont {Berti}\ \emph {et~al.}(2015)\citenamefont {Berti}
  \emph {et~al.}}]{Berti:2015itd}%
  \BibitemOpen
  \bibfield  {author} {\bibinfo {author} {\bibfnamefont {Emanuele}\
  \bibnamefont {Berti}} \emph {et~al.},\ }\bibfield  {title} {\enquote
  {\bibinfo {title} {{Testing General Relativity with Present and Future
  Astrophysical Observations}},}\ }\href {\doibase
  10.1088/0264-9381/32/24/243001} {\bibfield  {journal} {\bibinfo  {journal}
  {Class. Quant. Grav.}\ }\textbf {\bibinfo {volume} {32}},\ \bibinfo {pages}
  {243001} (\bibinfo {year} {2015})},\ \Eprint
  {http://arxiv.org/abs/1501.07274} {arXiv:1501.07274 [gr-qc]} \BibitemShut
  {NoStop}%
\bibitem [{\citenamefont {Barack}\ \emph {et~al.}(2019)\citenamefont {Barack}
  \emph {et~al.}}]{Barack:2018yly}%
  \BibitemOpen
  \bibfield  {author} {\bibinfo {author} {\bibfnamefont {Leor}\ \bibnamefont
  {Barack}} \emph {et~al.},\ }\bibfield  {title} {\enquote {\bibinfo {title}
  {{Black holes, gravitational waves and fundamental physics: a roadmap}},}\
  }\href {\doibase 10.1088/1361-6382/ab0587} {\bibfield  {journal} {\bibinfo
  {journal} {Class. Quant. Grav.}\ }\textbf {\bibinfo {volume} {36}},\ \bibinfo
  {pages} {143001} (\bibinfo {year} {2019})},\ \Eprint
  {http://arxiv.org/abs/1806.05195} {arXiv:1806.05195 [gr-qc]} \BibitemShut
  {NoStop}%
\bibitem [{\citenamefont {Yunes}\ \emph {et~al.}(2016)\citenamefont {Yunes},
  \citenamefont {Yagi},\ and\ \citenamefont {Pretorius}}]{Yunes:2016jcc}%
  \BibitemOpen
  \bibfield  {author} {\bibinfo {author} {\bibfnamefont {Nicolas}\ \bibnamefont
  {Yunes}}, \bibinfo {author} {\bibfnamefont {Kent}\ \bibnamefont {Yagi}}, \
  and\ \bibinfo {author} {\bibfnamefont {Frans}\ \bibnamefont {Pretorius}},\
  }\bibfield  {title} {\enquote {\bibinfo {title} {{Theoretical Physics
  Implications of the Binary Black-Hole Mergers GW150914 and GW151226}},}\
  }\href {\doibase 10.1103/PhysRevD.94.084002} {\bibfield  {journal} {\bibinfo
  {journal} {Phys. Rev. D}\ }\textbf {\bibinfo {volume} {94}},\ \bibinfo
  {pages} {084002} (\bibinfo {year} {2016})},\ \Eprint
  {http://arxiv.org/abs/1603.08955} {arXiv:1603.08955 [gr-qc]} \BibitemShut
  {NoStop}%
\bibitem [{\citenamefont {Cardoso}\ and\ \citenamefont
  {Gualtieri}(2016)}]{Cardoso:2016ryw}%
  \BibitemOpen
  \bibfield  {author} {\bibinfo {author} {\bibfnamefont {Vitor}\ \bibnamefont
  {Cardoso}}\ and\ \bibinfo {author} {\bibfnamefont {Leonardo}\ \bibnamefont
  {Gualtieri}},\ }\bibfield  {title} {\enquote {\bibinfo {title} {{Testing the
  black hole \textquoteleft{}no-hair\textquoteright{} hypothesis}},}\ }\href
  {\doibase 10.1088/0264-9381/33/17/174001} {\bibfield  {journal} {\bibinfo
  {journal} {Class. Quant. Grav.}\ }\textbf {\bibinfo {volume} {33}},\ \bibinfo
  {pages} {174001} (\bibinfo {year} {2016})},\ \Eprint
  {http://arxiv.org/abs/1607.03133} {arXiv:1607.03133 [gr-qc]} \BibitemShut
  {NoStop}%
\bibitem [{\citenamefont {Berti}\ \emph
  {et~al.}(2018{\natexlab{a}})\citenamefont {Berti}, \citenamefont {Yagi},\
  and\ \citenamefont {Yunes}}]{Berti:2018cxi}%
  \BibitemOpen
  \bibfield  {author} {\bibinfo {author} {\bibfnamefont {Emanuele}\
  \bibnamefont {Berti}}, \bibinfo {author} {\bibfnamefont {Kent}\ \bibnamefont
  {Yagi}}, \ and\ \bibinfo {author} {\bibfnamefont {Nicol\'as}\ \bibnamefont
  {Yunes}},\ }\bibfield  {title} {\enquote {\bibinfo {title} {{Extreme Gravity
  Tests with Gravitational Waves from Compact Binary Coalescences: (I)
  Inspiral-Merger}},}\ }\href {\doibase 10.1007/s10714-018-2362-8} {\bibfield
  {journal} {\bibinfo  {journal} {Gen. Rel. Grav.}\ }\textbf {\bibinfo {volume}
  {50}},\ \bibinfo {pages} {46} (\bibinfo {year} {2018}{\natexlab{a}})},\
  \Eprint {http://arxiv.org/abs/1801.03208} {arXiv:1801.03208 [gr-qc]}
  \BibitemShut {NoStop}%
\bibitem [{\citenamefont {Berti}\ \emph
  {et~al.}(2018{\natexlab{b}})\citenamefont {Berti}, \citenamefont {Yagi},
  \citenamefont {Yang},\ and\ \citenamefont {Yunes}}]{Berti:2018vdi}%
  \BibitemOpen
  \bibfield  {author} {\bibinfo {author} {\bibfnamefont {Emanuele}\
  \bibnamefont {Berti}}, \bibinfo {author} {\bibfnamefont {Kent}\ \bibnamefont
  {Yagi}}, \bibinfo {author} {\bibfnamefont {Huan}\ \bibnamefont {Yang}}, \
  and\ \bibinfo {author} {\bibfnamefont {Nicol\'as}\ \bibnamefont {Yunes}},\
  }\bibfield  {title} {\enquote {\bibinfo {title} {{Extreme Gravity Tests with
  Gravitational Waves from Compact Binary Coalescences: (II) Ringdown}},}\
  }\href {\doibase 10.1007/s10714-018-2372-6} {\bibfield  {journal} {\bibinfo
  {journal} {Gen. Rel. Grav.}\ }\textbf {\bibinfo {volume} {50}},\ \bibinfo
  {pages} {49} (\bibinfo {year} {2018}{\natexlab{b}})},\ \Eprint
  {http://arxiv.org/abs/1801.03587} {arXiv:1801.03587 [gr-qc]} \BibitemShut
  {NoStop}%
\bibitem [{\citenamefont {Cardoso}\ and\ \citenamefont
  {Pani}(2019)}]{Cardoso:2019rvt}%
  \BibitemOpen
  \bibfield  {author} {\bibinfo {author} {\bibfnamefont {Vitor}\ \bibnamefont
  {Cardoso}}\ and\ \bibinfo {author} {\bibfnamefont {Paolo}\ \bibnamefont
  {Pani}},\ }\bibfield  {title} {\enquote {\bibinfo {title} {{Testing the
  nature of dark compact objects: a status report}},}\ }\href {\doibase
  10.1007/s41114-019-0020-4} {\bibfield  {journal} {\bibinfo  {journal} {Living
  Rev. Rel.}\ }\textbf {\bibinfo {volume} {22}},\ \bibinfo {pages} {4}
  (\bibinfo {year} {2019})},\ \Eprint {http://arxiv.org/abs/1904.05363}
  {arXiv:1904.05363 [gr-qc]} \BibitemShut {NoStop}%
\bibitem [{\citenamefont {Abbott}\ \emph
  {et~al.}(2016{\natexlab{b}})\citenamefont {Abbott} \emph
  {et~al.}}]{LIGOScientific:2016lio}%
  \BibitemOpen
  \bibfield  {author} {\bibinfo {author} {\bibfnamefont {B.~P.}\ \bibnamefont
  {Abbott}} \emph {et~al.} (\bibinfo {collaboration} {LIGO Scientific,
  Virgo}),\ }\bibfield  {title} {\enquote {\bibinfo {title} {{Tests of general
  relativity with GW150914}},}\ }\href {\doibase
  10.1103/PhysRevLett.116.221101} {\bibfield  {journal} {\bibinfo  {journal}
  {Phys. Rev. Lett.}\ }\textbf {\bibinfo {volume} {116}},\ \bibinfo {pages}
  {221101} (\bibinfo {year} {2016}{\natexlab{b}})},\ \bibinfo {note} {[Erratum:
  Phys.Rev.Lett. 121, 129902 (2018)]},\ \Eprint
  {http://arxiv.org/abs/1602.03841} {arXiv:1602.03841 [gr-qc]} \BibitemShut
  {NoStop}%
\bibitem [{\citenamefont {Abbott}\ \emph
  {et~al.}(2019{\natexlab{b}})\citenamefont {Abbott} \emph
  {et~al.}}]{LIGOScientific:2019fpa}%
  \BibitemOpen
  \bibfield  {author} {\bibinfo {author} {\bibfnamefont {B.~P.}\ \bibnamefont
  {Abbott}} \emph {et~al.} (\bibinfo {collaboration} {LIGO Scientific,
  Virgo}),\ }\bibfield  {title} {\enquote {\bibinfo {title} {{Tests of General
  Relativity with the Binary Black Hole Signals from the LIGO-Virgo Catalog
  GWTC-1}},}\ }\href {\doibase 10.1103/PhysRevD.100.104036} {\bibfield
  {journal} {\bibinfo  {journal} {Phys. Rev. D}\ }\textbf {\bibinfo {volume}
  {100}},\ \bibinfo {pages} {104036} (\bibinfo {year} {2019}{\natexlab{b}})},\
  \Eprint {http://arxiv.org/abs/1903.04467} {arXiv:1903.04467 [gr-qc]}
  \BibitemShut {NoStop}%
\bibitem [{\citenamefont {Abbott}\ \emph
  {et~al.}(2021{\natexlab{c}})\citenamefont {Abbott} \emph
  {et~al.}}]{LIGOScientific:2020tif}%
  \BibitemOpen
  \bibfield  {author} {\bibinfo {author} {\bibfnamefont {R.}~\bibnamefont
  {Abbott}} \emph {et~al.} (\bibinfo {collaboration} {LIGO Scientific,
  Virgo}),\ }\bibfield  {title} {\enquote {\bibinfo {title} {{Tests of general
  relativity with binary black holes from the second LIGO-Virgo
  gravitational-wave transient catalog}},}\ }\href {\doibase
  10.1103/PhysRevD.103.122002} {\bibfield  {journal} {\bibinfo  {journal}
  {Phys. Rev. D}\ }\textbf {\bibinfo {volume} {103}},\ \bibinfo {pages}
  {122002} (\bibinfo {year} {2021}{\natexlab{c}})},\ \Eprint
  {http://arxiv.org/abs/2010.14529} {arXiv:2010.14529 [gr-qc]} \BibitemShut
  {NoStop}%
\bibitem [{\citenamefont {Abbott}\ \emph
  {et~al.}(2021{\natexlab{d}})\citenamefont {Abbott} \emph
  {et~al.}}]{LIGOScientific:2021sio}%
  \BibitemOpen
  \bibfield  {author} {\bibinfo {author} {\bibfnamefont {R.}~\bibnamefont
  {Abbott}} \emph {et~al.} (\bibinfo {collaboration} {LIGO Scientific, VIRGO,
  KAGRA}),\ }\bibfield  {title} {\enquote {\bibinfo {title} {{Tests of General
  Relativity with GWTC-3}},}\ }\href@noop {} {\  (\bibinfo {year}
  {2021}{\natexlab{d}})},\ \Eprint {http://arxiv.org/abs/2112.06861}
  {arXiv:2112.06861 [gr-qc]} \BibitemShut {NoStop}%
\bibitem [{\citenamefont {Ghosh}\ \emph {et~al.}(2021)\citenamefont {Ghosh},
  \citenamefont {Brito},\ and\ \citenamefont {Buonanno}}]{Ghosh:2021mrv}%
  \BibitemOpen
  \bibfield  {author} {\bibinfo {author} {\bibfnamefont {Abhirup}\ \bibnamefont
  {Ghosh}}, \bibinfo {author} {\bibfnamefont {Richard}\ \bibnamefont {Brito}},
  \ and\ \bibinfo {author} {\bibfnamefont {Alessandra}\ \bibnamefont
  {Buonanno}},\ }\bibfield  {title} {\enquote {\bibinfo {title} {{Constraints
  on quasinormal-mode frequencies with LIGO-Virgo binary\textendash{}black-hole
  observations}},}\ }\href {\doibase 10.1103/PhysRevD.103.124041} {\bibfield
  {journal} {\bibinfo  {journal} {Phys. Rev. D}\ }\textbf {\bibinfo {volume}
  {103}},\ \bibinfo {pages} {124041} (\bibinfo {year} {2021})},\ \Eprint
  {http://arxiv.org/abs/2104.01906} {arXiv:2104.01906 [gr-qc]} \BibitemShut
  {NoStop}%
\bibitem [{\citenamefont {Detweiler}(1980)}]{Detweiler:1980gk}%
  \BibitemOpen
  \bibfield  {author} {\bibinfo {author} {\bibfnamefont {Steven~L.}\
  \bibnamefont {Detweiler}},\ }\bibfield  {title} {\enquote {\bibinfo {title}
  {{Black holes and gravitational waves. III. The resonant frequencies of
  rotating holes}},}\ }\href {\doibase 10.1086/158109} {\bibfield  {journal}
  {\bibinfo  {journal} {Astrophys. J.}\ }\textbf {\bibinfo {volume} {239}},\
  \bibinfo {pages} {292--295} (\bibinfo {year} {1980})}\BibitemShut {NoStop}%
\bibitem [{\citenamefont {Dreyer}\ \emph {et~al.}(2004)\citenamefont {Dreyer},
  \citenamefont {Kelly}, \citenamefont {Krishnan}, \citenamefont {Finn},
  \citenamefont {Garrison},\ and\ \citenamefont
  {Lopez-Aleman}}]{Dreyer:2003bv}%
  \BibitemOpen
  \bibfield  {author} {\bibinfo {author} {\bibfnamefont {Olaf}\ \bibnamefont
  {Dreyer}}, \bibinfo {author} {\bibfnamefont {Bernard~J.}\ \bibnamefont
  {Kelly}}, \bibinfo {author} {\bibfnamefont {Badri}\ \bibnamefont {Krishnan}},
  \bibinfo {author} {\bibfnamefont {Lee~Samuel}\ \bibnamefont {Finn}}, \bibinfo
  {author} {\bibfnamefont {David}\ \bibnamefont {Garrison}}, \ and\ \bibinfo
  {author} {\bibfnamefont {Ramon}\ \bibnamefont {Lopez-Aleman}},\ }\bibfield
  {title} {\enquote {\bibinfo {title} {{Black hole spectroscopy: Testing
  general relativity through gravitational wave observations}},}\ }\href
  {\doibase 10.1088/0264-9381/21/4/003} {\bibfield  {journal} {\bibinfo
  {journal} {Class. Quant. Grav.}\ }\textbf {\bibinfo {volume} {21}},\ \bibinfo
  {pages} {787--804} (\bibinfo {year} {2004})},\ \Eprint
  {http://arxiv.org/abs/gr-qc/0309007} {arXiv:gr-qc/0309007} \BibitemShut
  {NoStop}%
\bibitem [{\citenamefont {Berti}\ \emph {et~al.}(2006)\citenamefont {Berti},
  \citenamefont {Cardoso},\ and\ \citenamefont {Will}}]{Berti:2005ys}%
  \BibitemOpen
  \bibfield  {author} {\bibinfo {author} {\bibfnamefont {Emanuele}\
  \bibnamefont {Berti}}, \bibinfo {author} {\bibfnamefont {Vitor}\ \bibnamefont
  {Cardoso}}, \ and\ \bibinfo {author} {\bibfnamefont {Clifford~M.}\
  \bibnamefont {Will}},\ }\bibfield  {title} {\enquote {\bibinfo {title} {{On
  gravitational-wave spectroscopy of massive black holes with the space
  interferometer LISA}},}\ }\href {\doibase 10.1103/PhysRevD.73.064030}
  {\bibfield  {journal} {\bibinfo  {journal} {Phys. Rev. D}\ }\textbf {\bibinfo
  {volume} {73}},\ \bibinfo {pages} {064030} (\bibinfo {year} {2006})},\
  \Eprint {http://arxiv.org/abs/gr-qc/0512160} {arXiv:gr-qc/0512160}
  \BibitemShut {NoStop}%
\bibitem [{\citenamefont {Isi}\ \emph {et~al.}(2019)\citenamefont {Isi},
  \citenamefont {Giesler}, \citenamefont {Farr}, \citenamefont {Scheel},\ and\
  \citenamefont {Teukolsky}}]{Isi:2019aib}%
  \BibitemOpen
  \bibfield  {author} {\bibinfo {author} {\bibfnamefont {Maximiliano}\
  \bibnamefont {Isi}}, \bibinfo {author} {\bibfnamefont {Matthew}\ \bibnamefont
  {Giesler}}, \bibinfo {author} {\bibfnamefont {Will~M.}\ \bibnamefont {Farr}},
  \bibinfo {author} {\bibfnamefont {Mark~A.}\ \bibnamefont {Scheel}}, \ and\
  \bibinfo {author} {\bibfnamefont {Saul~A.}\ \bibnamefont {Teukolsky}},\
  }\bibfield  {title} {\enquote {\bibinfo {title} {{Testing the no-hair theorem
  with GW150914}},}\ }\href {\doibase 10.1103/PhysRevLett.123.111102}
  {\bibfield  {journal} {\bibinfo  {journal} {Phys. Rev. Lett.}\ }\textbf
  {\bibinfo {volume} {123}},\ \bibinfo {pages} {111102} (\bibinfo {year}
  {2019})},\ \Eprint {http://arxiv.org/abs/1905.00869} {arXiv:1905.00869
  [gr-qc]} \BibitemShut {NoStop}%
\bibitem [{\citenamefont {Capano}\ \emph {et~al.}(2021)\citenamefont {Capano},
  \citenamefont {Cabero}, \citenamefont {Westerweck}, \citenamefont {Abedi},
  \citenamefont {Kastha}, \citenamefont {Nitz}, \citenamefont {Nielsen},\ and\
  \citenamefont {Krishnan}}]{Capano:2021etf}%
  \BibitemOpen
  \bibfield  {author} {\bibinfo {author} {\bibfnamefont {Collin~D.}\
  \bibnamefont {Capano}}, \bibinfo {author} {\bibfnamefont {Miriam}\
  \bibnamefont {Cabero}}, \bibinfo {author} {\bibfnamefont {Julian}\
  \bibnamefont {Westerweck}}, \bibinfo {author} {\bibfnamefont {Jahed}\
  \bibnamefont {Abedi}}, \bibinfo {author} {\bibfnamefont {Shilpa}\
  \bibnamefont {Kastha}}, \bibinfo {author} {\bibfnamefont {Alexander~H.}\
  \bibnamefont {Nitz}}, \bibinfo {author} {\bibfnamefont {Alex~B.}\
  \bibnamefont {Nielsen}}, \ and\ \bibinfo {author} {\bibfnamefont {Badri}\
  \bibnamefont {Krishnan}},\ }\bibfield  {title} {\enquote {\bibinfo {title}
  {{Observation of a multimode quasi-normal spectrum from a perturbed black
  hole}},}\ }\href@noop {} {\  (\bibinfo {year} {2021})},\ \Eprint
  {http://arxiv.org/abs/2105.05238} {arXiv:2105.05238 [gr-qc]} \BibitemShut
  {NoStop}%
\bibitem [{\citenamefont {Cotesta}\ \emph {et~al.}(2022)\citenamefont
  {Cotesta}, \citenamefont {Carullo}, \citenamefont {Berti},\ and\
  \citenamefont {Cardoso}}]{Cotesta:2022pci}%
  \BibitemOpen
  \bibfield  {author} {\bibinfo {author} {\bibfnamefont {Roberto}\ \bibnamefont
  {Cotesta}}, \bibinfo {author} {\bibfnamefont {Gregorio}\ \bibnamefont
  {Carullo}}, \bibinfo {author} {\bibfnamefont {Emanuele}\ \bibnamefont
  {Berti}}, \ and\ \bibinfo {author} {\bibfnamefont {Vitor}\ \bibnamefont
  {Cardoso}},\ }\bibfield  {title} {\enquote {\bibinfo {title} {{Analysis of
  Ringdown Overtones in GW150914}},}\ }\href {\doibase
  10.1103/PhysRevLett.129.111102} {\bibfield  {journal} {\bibinfo  {journal}
  {Phys. Rev. Lett.}\ }\textbf {\bibinfo {volume} {129}},\ \bibinfo {pages}
  {111102} (\bibinfo {year} {2022})},\ \Eprint
  {http://arxiv.org/abs/2201.00822} {arXiv:2201.00822 [gr-qc]} \BibitemShut
  {NoStop}%
\bibitem [{\citenamefont {Finch}\ and\ \citenamefont
  {Moore}(2022)}]{Finch:2022ynt}%
  \BibitemOpen
  \bibfield  {author} {\bibinfo {author} {\bibfnamefont {Eliot}\ \bibnamefont
  {Finch}}\ and\ \bibinfo {author} {\bibfnamefont {Christopher~J.}\
  \bibnamefont {Moore}},\ }\bibfield  {title} {\enquote {\bibinfo {title}
  {{Searching for a ringdown overtone in GW150914}},}\ }\href {\doibase
  10.1103/PhysRevD.106.043005} {\bibfield  {journal} {\bibinfo  {journal}
  {Phys. Rev. D}\ }\textbf {\bibinfo {volume} {106}},\ \bibinfo {pages}
  {043005} (\bibinfo {year} {2022})},\ \Eprint
  {http://arxiv.org/abs/2205.07809} {arXiv:2205.07809 [gr-qc]} \BibitemShut
  {NoStop}%
\bibitem [{\citenamefont {Isi}\ and\ \citenamefont {Farr}(2022)}]{Isi:2022mhy}%
  \BibitemOpen
  \bibfield  {author} {\bibinfo {author} {\bibfnamefont {Maximiliano}\
  \bibnamefont {Isi}}\ and\ \bibinfo {author} {\bibfnamefont {Will~M.}\
  \bibnamefont {Farr}},\ }\bibfield  {title} {\enquote {\bibinfo {title}
  {{Revisiting the ringdown of GW150914}},}\ }\href@noop {} {\  (\bibinfo
  {year} {2022})},\ \Eprint {http://arxiv.org/abs/2202.02941} {arXiv:2202.02941
  [gr-qc]} \BibitemShut {NoStop}%
\bibitem [{\citenamefont {Capano}\ \emph {et~al.}(2022)\citenamefont {Capano},
  \citenamefont {Abedi}, \citenamefont {Kastha}, \citenamefont {Nitz},
  \citenamefont {Westerweck}, \citenamefont {Wang}, \citenamefont {Cabero},
  \citenamefont {Nielsen},\ and\ \citenamefont {Krishnan}}]{Capano:2022zqm}%
  \BibitemOpen
  \bibfield  {author} {\bibinfo {author} {\bibfnamefont {Collin~D.}\
  \bibnamefont {Capano}}, \bibinfo {author} {\bibfnamefont {Jahed}\
  \bibnamefont {Abedi}}, \bibinfo {author} {\bibfnamefont {Shilpa}\
  \bibnamefont {Kastha}}, \bibinfo {author} {\bibfnamefont {Alexander~H.}\
  \bibnamefont {Nitz}}, \bibinfo {author} {\bibfnamefont {Julian}\ \bibnamefont
  {Westerweck}}, \bibinfo {author} {\bibfnamefont {Yi-Fan}\ \bibnamefont
  {Wang}}, \bibinfo {author} {\bibfnamefont {Miriam}\ \bibnamefont {Cabero}},
  \bibinfo {author} {\bibfnamefont {Alex~B.}\ \bibnamefont {Nielsen}}, \ and\
  \bibinfo {author} {\bibfnamefont {Badri}\ \bibnamefont {Krishnan}},\
  }\bibfield  {title} {\enquote {\bibinfo {title} {{Statistical validation of
  the detection of a sub-dominant quasi-normal mode in GW190521}},}\
  }\href@noop {} {\  (\bibinfo {year} {2022})},\ \Eprint
  {http://arxiv.org/abs/2209.00640} {arXiv:2209.00640 [gr-qc]} \BibitemShut
  {NoStop}%
\bibitem [{\citenamefont {Sberna}\ \emph {et~al.}(2022)\citenamefont {Sberna},
  \citenamefont {Bosch}, \citenamefont {East}, \citenamefont {Green},\ and\
  \citenamefont {Lehner}}]{Sberna:2021eui}%
  \BibitemOpen
  \bibfield  {author} {\bibinfo {author} {\bibfnamefont {Laura}\ \bibnamefont
  {Sberna}}, \bibinfo {author} {\bibfnamefont {Pablo}\ \bibnamefont {Bosch}},
  \bibinfo {author} {\bibfnamefont {William~E.}\ \bibnamefont {East}}, \bibinfo
  {author} {\bibfnamefont {Stephen~R.}\ \bibnamefont {Green}}, \ and\ \bibinfo
  {author} {\bibfnamefont {Luis}\ \bibnamefont {Lehner}},\ }\bibfield  {title}
  {\enquote {\bibinfo {title} {{Nonlinear effects in the black hole ringdown:
  Absorption-induced mode excitation}},}\ }\href {\doibase
  10.1103/PhysRevD.105.064046} {\bibfield  {journal} {\bibinfo  {journal}
  {Phys. Rev. D}\ }\textbf {\bibinfo {volume} {105}},\ \bibinfo {pages}
  {064046} (\bibinfo {year} {2022})},\ \Eprint
  {http://arxiv.org/abs/2112.11168} {arXiv:2112.11168 [gr-qc]} \BibitemShut
  {NoStop}%
\bibitem [{\citenamefont {Cheung}\ \emph {et~al.}(2022)\citenamefont {Cheung}
  \emph {et~al.}}]{Cheung:2022rbm}%
  \BibitemOpen
  \bibfield  {author} {\bibinfo {author} {\bibfnamefont {Mark Ho-Yeuk}\
  \bibnamefont {Cheung}} \emph {et~al.},\ }\bibfield  {title} {\enquote
  {\bibinfo {title} {{Nonlinear effects in black hole ringdown}},}\ }\href@noop
  {} {\  (\bibinfo {year} {2022})},\ \Eprint {http://arxiv.org/abs/2208.07374}
  {arXiv:2208.07374 [gr-qc]} \BibitemShut {NoStop}%
\bibitem [{\citenamefont {Mitman}\ \emph {et~al.}(2022)\citenamefont {Mitman}
  \emph {et~al.}}]{Mitman:2022qdl}%
  \BibitemOpen
  \bibfield  {author} {\bibinfo {author} {\bibfnamefont {Keefe}\ \bibnamefont
  {Mitman}} \emph {et~al.},\ }\bibfield  {title} {\enquote {\bibinfo {title}
  {{Nonlinearities in black hole ringdowns}},}\ }\href@noop {} {\  (\bibinfo
  {year} {2022})},\ \Eprint {http://arxiv.org/abs/2208.07380} {arXiv:2208.07380
  [gr-qc]} \BibitemShut {NoStop}%
\bibitem [{\citenamefont {Berti}\ \emph {et~al.}(2007)\citenamefont {Berti},
  \citenamefont {Cardoso}, \citenamefont {Cardoso},\ and\ \citenamefont
  {Cavaglia}}]{Berti:2007zu}%
  \BibitemOpen
  \bibfield  {author} {\bibinfo {author} {\bibfnamefont {Emanuele}\
  \bibnamefont {Berti}}, \bibinfo {author} {\bibfnamefont {Jaime}\ \bibnamefont
  {Cardoso}}, \bibinfo {author} {\bibfnamefont {Vitor}\ \bibnamefont
  {Cardoso}}, \ and\ \bibinfo {author} {\bibfnamefont {Marco}\ \bibnamefont
  {Cavaglia}},\ }\bibfield  {title} {\enquote {\bibinfo {title}
  {{Matched-filtering and parameter estimation of ringdown waveforms}},}\
  }\href {\doibase 10.1103/PhysRevD.76.104044} {\bibfield  {journal} {\bibinfo
  {journal} {Phys. Rev. D}\ }\textbf {\bibinfo {volume} {76}},\ \bibinfo
  {pages} {104044} (\bibinfo {year} {2007})},\ \Eprint
  {http://arxiv.org/abs/0707.1202} {arXiv:0707.1202 [gr-qc]} \BibitemShut
  {NoStop}%
\bibitem [{\citenamefont {Baibhav}\ \emph {et~al.}(2018)\citenamefont
  {Baibhav}, \citenamefont {Berti}, \citenamefont {Cardoso},\ and\
  \citenamefont {Khanna}}]{Baibhav:2017jhs}%
  \BibitemOpen
  \bibfield  {author} {\bibinfo {author} {\bibfnamefont {Vishal}\ \bibnamefont
  {Baibhav}}, \bibinfo {author} {\bibfnamefont {Emanuele}\ \bibnamefont
  {Berti}}, \bibinfo {author} {\bibfnamefont {Vitor}\ \bibnamefont {Cardoso}},
  \ and\ \bibinfo {author} {\bibfnamefont {Gaurav}\ \bibnamefont {Khanna}},\
  }\bibfield  {title} {\enquote {\bibinfo {title} {{Black Hole Spectroscopy:
  Systematic Errors and Ringdown Energy Estimates}},}\ }\href {\doibase
  10.1103/PhysRevD.97.044048} {\bibfield  {journal} {\bibinfo  {journal} {Phys.
  Rev. D}\ }\textbf {\bibinfo {volume} {97}},\ \bibinfo {pages} {044048}
  (\bibinfo {year} {2018})},\ \Eprint {http://arxiv.org/abs/1710.02156}
  {arXiv:1710.02156 [gr-qc]} \BibitemShut {NoStop}%
\bibitem [{\citenamefont {Giesler}\ \emph {et~al.}(2019)\citenamefont
  {Giesler}, \citenamefont {Isi}, \citenamefont {Scheel},\ and\ \citenamefont
  {Teukolsky}}]{Giesler:2019uxc}%
  \BibitemOpen
  \bibfield  {author} {\bibinfo {author} {\bibfnamefont {Matthew}\ \bibnamefont
  {Giesler}}, \bibinfo {author} {\bibfnamefont {Maximiliano}\ \bibnamefont
  {Isi}}, \bibinfo {author} {\bibfnamefont {Mark~A.}\ \bibnamefont {Scheel}}, \
  and\ \bibinfo {author} {\bibfnamefont {Saul}\ \bibnamefont {Teukolsky}},\
  }\bibfield  {title} {\enquote {\bibinfo {title} {{Black Hole Ringdown: The
  Importance of Overtones}},}\ }\href {\doibase 10.1103/PhysRevX.9.041060}
  {\bibfield  {journal} {\bibinfo  {journal} {Phys. Rev. X}\ }\textbf {\bibinfo
  {volume} {9}},\ \bibinfo {pages} {041060} (\bibinfo {year} {2019})},\ \Eprint
  {http://arxiv.org/abs/1903.08284} {arXiv:1903.08284 [gr-qc]} \BibitemShut
  {NoStop}%
\bibitem [{\citenamefont {Ota}\ and\ \citenamefont
  {Chirenti}(2020)}]{Ota:2019bzl}%
  \BibitemOpen
  \bibfield  {author} {\bibinfo {author} {\bibfnamefont {Iara}\ \bibnamefont
  {Ota}}\ and\ \bibinfo {author} {\bibfnamefont {Cecilia}\ \bibnamefont
  {Chirenti}},\ }\bibfield  {title} {\enquote {\bibinfo {title} {{Overtones or
  higher harmonics? Prospects for testing the no-hair theorem with
  gravitational wave detections}},}\ }\href {\doibase
  10.1103/PhysRevD.101.104005} {\bibfield  {journal} {\bibinfo  {journal}
  {Phys. Rev. D}\ }\textbf {\bibinfo {volume} {101}},\ \bibinfo {pages}
  {104005} (\bibinfo {year} {2020})},\ \Eprint
  {http://arxiv.org/abs/1911.00440} {arXiv:1911.00440 [gr-qc]} \BibitemShut
  {NoStop}%
\bibitem [{\citenamefont {Bhagwat}\ \emph
  {et~al.}(2020{\natexlab{a}})\citenamefont {Bhagwat}, \citenamefont {Forteza},
  \citenamefont {Pani},\ and\ \citenamefont {Ferrari}}]{Bhagwat:2019dtm}%
  \BibitemOpen
  \bibfield  {author} {\bibinfo {author} {\bibfnamefont {Swetha}\ \bibnamefont
  {Bhagwat}}, \bibinfo {author} {\bibfnamefont {Xisco~Jimenez}\ \bibnamefont
  {Forteza}}, \bibinfo {author} {\bibfnamefont {Paolo}\ \bibnamefont {Pani}}, \
  and\ \bibinfo {author} {\bibfnamefont {Valeria}\ \bibnamefont {Ferrari}},\
  }\bibfield  {title} {\enquote {\bibinfo {title} {{Ringdown overtones, black
  hole spectroscopy, and no-hair theorem tests}},}\ }\href {\doibase
  10.1103/PhysRevD.101.044033} {\bibfield  {journal} {\bibinfo  {journal}
  {Phys. Rev. D}\ }\textbf {\bibinfo {volume} {101}},\ \bibinfo {pages}
  {044033} (\bibinfo {year} {2020}{\natexlab{a}})},\ \Eprint
  {http://arxiv.org/abs/1910.08708} {arXiv:1910.08708 [gr-qc]} \BibitemShut
  {NoStop}%
\bibitem [{\citenamefont {Bhagwat}\ \emph
  {et~al.}(2020{\natexlab{b}})\citenamefont {Bhagwat}, \citenamefont {Cabero},
  \citenamefont {Capano}, \citenamefont {Krishnan},\ and\ \citenamefont
  {Brown}}]{Bhagwat:2019bwv}%
  \BibitemOpen
  \bibfield  {author} {\bibinfo {author} {\bibfnamefont {Swetha}\ \bibnamefont
  {Bhagwat}}, \bibinfo {author} {\bibfnamefont {Miriam}\ \bibnamefont
  {Cabero}}, \bibinfo {author} {\bibfnamefont {Collin~D.}\ \bibnamefont
  {Capano}}, \bibinfo {author} {\bibfnamefont {Badri}\ \bibnamefont
  {Krishnan}}, \ and\ \bibinfo {author} {\bibfnamefont {Duncan~A.}\
  \bibnamefont {Brown}},\ }\bibfield  {title} {\enquote {\bibinfo {title}
  {{Detectability of the subdominant mode in a binary black hole ringdown}},}\
  }\href {\doibase 10.1103/PhysRevD.102.024023} {\bibfield  {journal} {\bibinfo
   {journal} {Phys. Rev. D}\ }\textbf {\bibinfo {volume} {102}},\ \bibinfo
  {pages} {024023} (\bibinfo {year} {2020}{\natexlab{b}})},\ \Eprint
  {http://arxiv.org/abs/1910.13203} {arXiv:1910.13203 [gr-qc]} \BibitemShut
  {NoStop}%
\bibitem [{\citenamefont {Cook}(2020)}]{Cook:2020otn}%
  \BibitemOpen
  \bibfield  {author} {\bibinfo {author} {\bibfnamefont {Gregory~B.}\
  \bibnamefont {Cook}},\ }\bibfield  {title} {\enquote {\bibinfo {title}
  {{Aspects of multimode Kerr ringdown fitting}},}\ }\href {\doibase
  10.1103/PhysRevD.102.024027} {\bibfield  {journal} {\bibinfo  {journal}
  {Phys. Rev. D}\ }\textbf {\bibinfo {volume} {102}},\ \bibinfo {pages}
  {024027} (\bibinfo {year} {2020})},\ \Eprint
  {http://arxiv.org/abs/2004.08347} {arXiv:2004.08347 [gr-qc]} \BibitemShut
  {NoStop}%
\bibitem [{\citenamefont {Bustillo}\ \emph {et~al.}(2021)\citenamefont
  {Bustillo}, \citenamefont {Lasky},\ and\ \citenamefont
  {Thrane}}]{Bustillo:2020buq}%
  \BibitemOpen
  \bibfield  {author} {\bibinfo {author} {\bibfnamefont {Juan~Calder\'on}\
  \bibnamefont {Bustillo}}, \bibinfo {author} {\bibfnamefont {Paul~D.}\
  \bibnamefont {Lasky}}, \ and\ \bibinfo {author} {\bibfnamefont {Eric}\
  \bibnamefont {Thrane}},\ }\bibfield  {title} {\enquote {\bibinfo {title}
  {{Black-hole spectroscopy, the no-hair theorem, and GW150914: Kerr versus
  Occam}},}\ }\href {\doibase 10.1103/PhysRevD.103.024041} {\bibfield
  {journal} {\bibinfo  {journal} {Phys. Rev. D}\ }\textbf {\bibinfo {volume}
  {103}},\ \bibinfo {pages} {024041} (\bibinfo {year} {2021})},\ \Eprint
  {http://arxiv.org/abs/2010.01857} {arXiv:2010.01857 [gr-qc]} \BibitemShut
  {NoStop}%
\bibitem [{\citenamefont {Jim\'enez~Forteza}\ \emph {et~al.}(2020)\citenamefont
  {Jim\'enez~Forteza}, \citenamefont {Bhagwat}, \citenamefont {Pani},\ and\
  \citenamefont {Ferrari}}]{Forteza:2020hbw}%
  \BibitemOpen
  \bibfield  {author} {\bibinfo {author} {\bibfnamefont {Xisco}\ \bibnamefont
  {Jim\'enez~Forteza}}, \bibinfo {author} {\bibfnamefont {Swetha}\ \bibnamefont
  {Bhagwat}}, \bibinfo {author} {\bibfnamefont {Paolo}\ \bibnamefont {Pani}}, \
  and\ \bibinfo {author} {\bibfnamefont {Valeria}\ \bibnamefont {Ferrari}},\
  }\bibfield  {title} {\enquote {\bibinfo {title} {{Spectroscopy of binary
  black hole ringdown using overtones and angular modes}},}\ }\href {\doibase
  10.1103/PhysRevD.102.044053} {\bibfield  {journal} {\bibinfo  {journal}
  {Phys. Rev. D}\ }\textbf {\bibinfo {volume} {102}},\ \bibinfo {pages}
  {044053} (\bibinfo {year} {2020})},\ \Eprint
  {http://arxiv.org/abs/2005.03260} {arXiv:2005.03260 [gr-qc]} \BibitemShut
  {NoStop}%
\bibitem [{\citenamefont {Maga\~na Zertuche}\ \emph {et~al.}(2022)\citenamefont
  {Maga\~na Zertuche} \emph {et~al.}}]{MaganaZertuche:2021syq}%
  \BibitemOpen
  \bibfield  {author} {\bibinfo {author} {\bibfnamefont {Lorena}\ \bibnamefont
  {Maga\~na Zertuche}} \emph {et~al.},\ }\bibfield  {title} {\enquote {\bibinfo
  {title} {{High Precision Ringdown Modeling: Multimode Fits and BMS
  Frames}},}\ }\href {\doibase 10.1103/PhysRevD.105.104015} {\bibfield
  {journal} {\bibinfo  {journal} {Phys. Rev. D}\ }\textbf {\bibinfo {volume}
  {105}},\ \bibinfo {pages} {104015} (\bibinfo {year} {2022})},\ \Eprint
  {http://arxiv.org/abs/2110.15922} {arXiv:2110.15922 [gr-qc]} \BibitemShut
  {NoStop}%
\bibitem [{\citenamefont {Berti}\ \emph {et~al.}(2016)\citenamefont {Berti},
  \citenamefont {Sesana}, \citenamefont {Barausse}, \citenamefont {Cardoso},\
  and\ \citenamefont {Belczynski}}]{Berti:2016lat}%
  \BibitemOpen
  \bibfield  {author} {\bibinfo {author} {\bibfnamefont {Emanuele}\
  \bibnamefont {Berti}}, \bibinfo {author} {\bibfnamefont {Alberto}\
  \bibnamefont {Sesana}}, \bibinfo {author} {\bibfnamefont {Enrico}\
  \bibnamefont {Barausse}}, \bibinfo {author} {\bibfnamefont {Vitor}\
  \bibnamefont {Cardoso}}, \ and\ \bibinfo {author} {\bibfnamefont {Krzysztof}\
  \bibnamefont {Belczynski}},\ }\bibfield  {title} {\enquote {\bibinfo {title}
  {{Spectroscopy of Kerr black holes with Earth- and space-based
  interferometers}},}\ }\href {\doibase 10.1103/PhysRevLett.117.101102}
  {\bibfield  {journal} {\bibinfo  {journal} {Phys. Rev. Lett.}\ }\textbf
  {\bibinfo {volume} {117}},\ \bibinfo {pages} {101102} (\bibinfo {year}
  {2016})},\ \Eprint {http://arxiv.org/abs/1605.09286} {arXiv:1605.09286
  [gr-qc]} \BibitemShut {NoStop}%
\bibitem [{\citenamefont {Cabero}\ \emph {et~al.}(2020)\citenamefont {Cabero},
  \citenamefont {Westerweck}, \citenamefont {Capano}, \citenamefont {Kumar},
  \citenamefont {Nielsen},\ and\ \citenamefont {Krishnan}}]{Cabero:2019zyt}%
  \BibitemOpen
  \bibfield  {author} {\bibinfo {author} {\bibfnamefont {Miriam}\ \bibnamefont
  {Cabero}}, \bibinfo {author} {\bibfnamefont {Julian}\ \bibnamefont
  {Westerweck}}, \bibinfo {author} {\bibfnamefont {Collin~D.}\ \bibnamefont
  {Capano}}, \bibinfo {author} {\bibfnamefont {Sumit}\ \bibnamefont {Kumar}},
  \bibinfo {author} {\bibfnamefont {Alex~B.}\ \bibnamefont {Nielsen}}, \ and\
  \bibinfo {author} {\bibfnamefont {Badri}\ \bibnamefont {Krishnan}},\
  }\bibfield  {title} {\enquote {\bibinfo {title} {{Black hole spectroscopy in
  the next decade}},}\ }\href {\doibase 10.1103/PhysRevD.101.064044} {\bibfield
   {journal} {\bibinfo  {journal} {Phys. Rev. D}\ }\textbf {\bibinfo {volume}
  {101}},\ \bibinfo {pages} {064044} (\bibinfo {year} {2020})},\ \Eprint
  {http://arxiv.org/abs/1911.01361} {arXiv:1911.01361 [gr-qc]} \BibitemShut
  {NoStop}%
\bibitem [{\citenamefont {Ota}\ and\ \citenamefont
  {Chirenti}(2022)}]{Ota:2021ypb}%
  \BibitemOpen
  \bibfield  {author} {\bibinfo {author} {\bibfnamefont {Iara}\ \bibnamefont
  {Ota}}\ and\ \bibinfo {author} {\bibfnamefont {Cecilia}\ \bibnamefont
  {Chirenti}},\ }\bibfield  {title} {\enquote {\bibinfo {title} {{Black hole
  spectroscopy horizons for current and future gravitational wave
  detectors}},}\ }\href {\doibase 10.1103/PhysRevD.105.044015} {\bibfield
  {journal} {\bibinfo  {journal} {Phys. Rev. D}\ }\textbf {\bibinfo {volume}
  {105}},\ \bibinfo {pages} {044015} (\bibinfo {year} {2022})},\ \Eprint
  {http://arxiv.org/abs/2108.01774} {arXiv:2108.01774 [gr-qc]} \BibitemShut
  {NoStop}%
\bibitem [{\citenamefont {Bhagwat}\ \emph {et~al.}(2022)\citenamefont
  {Bhagwat}, \citenamefont {Pacilio}, \citenamefont {Barausse},\ and\
  \citenamefont {Pani}}]{Bhagwat:2021kwv}%
  \BibitemOpen
  \bibfield  {author} {\bibinfo {author} {\bibfnamefont {Swetha}\ \bibnamefont
  {Bhagwat}}, \bibinfo {author} {\bibfnamefont {Costantino}\ \bibnamefont
  {Pacilio}}, \bibinfo {author} {\bibfnamefont {Enrico}\ \bibnamefont
  {Barausse}}, \ and\ \bibinfo {author} {\bibfnamefont {Paolo}\ \bibnamefont
  {Pani}},\ }\bibfield  {title} {\enquote {\bibinfo {title} {{Landscape of
  massive black-hole spectroscopy with LISA and the Einstein Telescope}},}\
  }\href {\doibase 10.1103/PhysRevD.105.124063} {\bibfield  {journal} {\bibinfo
   {journal} {Phys. Rev. D}\ }\textbf {\bibinfo {volume} {105}},\ \bibinfo
  {pages} {124063} (\bibinfo {year} {2022})},\ \Eprint
  {http://arxiv.org/abs/2201.00023} {arXiv:2201.00023 [gr-qc]} \BibitemShut
  {NoStop}%
\bibitem [{\citenamefont {Tattersall}\ \emph {et~al.}(2018)\citenamefont
  {Tattersall}, \citenamefont {Ferreira},\ and\ \citenamefont
  {Lagos}}]{Tattersall:2017erk}%
  \BibitemOpen
  \bibfield  {author} {\bibinfo {author} {\bibfnamefont {Oliver~J.}\
  \bibnamefont {Tattersall}}, \bibinfo {author} {\bibfnamefont {Pedro~G.}\
  \bibnamefont {Ferreira}}, \ and\ \bibinfo {author} {\bibfnamefont {Macarena}\
  \bibnamefont {Lagos}},\ }\bibfield  {title} {\enquote {\bibinfo {title}
  {{General theories of linear gravitational perturbations to a Schwarzschild
  Black Hole}},}\ }\href {\doibase 10.1103/PhysRevD.97.044021} {\bibfield
  {journal} {\bibinfo  {journal} {Phys. Rev. D}\ }\textbf {\bibinfo {volume}
  {97}},\ \bibinfo {pages} {044021} (\bibinfo {year} {2018})},\ \Eprint
  {http://arxiv.org/abs/1711.01992} {arXiv:1711.01992 [gr-qc]} \BibitemShut
  {NoStop}%
\bibitem [{\citenamefont {Cardoso}\ \emph {et~al.}(2019)\citenamefont
  {Cardoso}, \citenamefont {Kimura}, \citenamefont {Maselli}, \citenamefont
  {Berti}, \citenamefont {Macedo},\ and\ \citenamefont
  {McManus}}]{Cardoso:2019mqo}%
  \BibitemOpen
  \bibfield  {author} {\bibinfo {author} {\bibfnamefont {Vitor}\ \bibnamefont
  {Cardoso}}, \bibinfo {author} {\bibfnamefont {Masashi}\ \bibnamefont
  {Kimura}}, \bibinfo {author} {\bibfnamefont {Andrea}\ \bibnamefont
  {Maselli}}, \bibinfo {author} {\bibfnamefont {Emanuele}\ \bibnamefont
  {Berti}}, \bibinfo {author} {\bibfnamefont {Caio F.~B.}\ \bibnamefont
  {Macedo}}, \ and\ \bibinfo {author} {\bibfnamefont {Ryan}\ \bibnamefont
  {McManus}},\ }\bibfield  {title} {\enquote {\bibinfo {title} {{Parametrized
  black hole quasinormal ringdown: Decoupled equations for nonrotating black
  holes}},}\ }\href {\doibase 10.1103/PhysRevD.99.104077} {\bibfield  {journal}
  {\bibinfo  {journal} {Phys. Rev. D}\ }\textbf {\bibinfo {volume} {99}},\
  \bibinfo {pages} {104077} (\bibinfo {year} {2019})},\ \Eprint
  {http://arxiv.org/abs/1901.01265} {arXiv:1901.01265 [gr-qc]} \BibitemShut
  {NoStop}%
\bibitem [{\citenamefont {McManus}\ \emph {et~al.}(2019)\citenamefont
  {McManus}, \citenamefont {Berti}, \citenamefont {Macedo}, \citenamefont
  {Kimura}, \citenamefont {Maselli},\ and\ \citenamefont
  {Cardoso}}]{McManus:2019ulj}%
  \BibitemOpen
  \bibfield  {author} {\bibinfo {author} {\bibfnamefont {Ryan}\ \bibnamefont
  {McManus}}, \bibinfo {author} {\bibfnamefont {Emanuele}\ \bibnamefont
  {Berti}}, \bibinfo {author} {\bibfnamefont {Caio F.~B.}\ \bibnamefont
  {Macedo}}, \bibinfo {author} {\bibfnamefont {Masashi}\ \bibnamefont
  {Kimura}}, \bibinfo {author} {\bibfnamefont {Andrea}\ \bibnamefont
  {Maselli}}, \ and\ \bibinfo {author} {\bibfnamefont {Vitor}\ \bibnamefont
  {Cardoso}},\ }\bibfield  {title} {\enquote {\bibinfo {title} {{Parametrized
  black hole quasinormal ringdown. II. Coupled equations and quadratic
  corrections for nonrotating black holes}},}\ }\href {\doibase
  10.1103/PhysRevD.100.044061} {\bibfield  {journal} {\bibinfo  {journal}
  {Phys. Rev. D}\ }\textbf {\bibinfo {volume} {100}},\ \bibinfo {pages}
  {044061} (\bibinfo {year} {2019})},\ \Eprint
  {http://arxiv.org/abs/1906.05155} {arXiv:1906.05155 [gr-qc]} \BibitemShut
  {NoStop}%
\bibitem [{\citenamefont {V\"olkel}\ \emph {et~al.}(2022)\citenamefont
  {V\"olkel}, \citenamefont {Franchini},\ and\ \citenamefont
  {Barausse}}]{Volkel:2022aca}%
  \BibitemOpen
  \bibfield  {author} {\bibinfo {author} {\bibfnamefont {Sebastian~H.}\
  \bibnamefont {V\"olkel}}, \bibinfo {author} {\bibfnamefont {Nicola}\
  \bibnamefont {Franchini}}, \ and\ \bibinfo {author} {\bibfnamefont {Enrico}\
  \bibnamefont {Barausse}},\ }\bibfield  {title} {\enquote {\bibinfo {title}
  {{Theory-agnostic reconstruction of potential and couplings from quasinormal
  modes}},}\ }\href {\doibase 10.1103/PhysRevD.105.084046} {\bibfield
  {journal} {\bibinfo  {journal} {Phys. Rev. D}\ }\textbf {\bibinfo {volume}
  {105}},\ \bibinfo {pages} {084046} (\bibinfo {year} {2022})},\ \Eprint
  {http://arxiv.org/abs/2202.08655} {arXiv:2202.08655 [gr-qc]} \BibitemShut
  {NoStop}%
\bibitem [{\citenamefont {V\"olkel}\ and\ \citenamefont
  {Barausse}(2020)}]{Volkel:2020daa}%
  \BibitemOpen
  \bibfield  {author} {\bibinfo {author} {\bibfnamefont {Sebastian~H.}\
  \bibnamefont {V\"olkel}}\ and\ \bibinfo {author} {\bibfnamefont {Enrico}\
  \bibnamefont {Barausse}},\ }\bibfield  {title} {\enquote {\bibinfo {title}
  {{Bayesian Metric Reconstruction with Gravitational Wave Observations}},}\
  }\href {\doibase 10.1103/PhysRevD.102.084025} {\bibfield  {journal} {\bibinfo
   {journal} {Phys. Rev. D}\ }\textbf {\bibinfo {volume} {102}},\ \bibinfo
  {pages} {084025} (\bibinfo {year} {2020})},\ \Eprint
  {http://arxiv.org/abs/2007.02986} {arXiv:2007.02986 [gr-qc]} \BibitemShut
  {NoStop}%
\bibitem [{\citenamefont {Konoplya}\ and\ \citenamefont
  {Zhidenko}(2022)}]{Konoplya:2022pbc}%
  \BibitemOpen
  \bibfield  {author} {\bibinfo {author} {\bibfnamefont {R.~A.}\ \bibnamefont
  {Konoplya}}\ and\ \bibinfo {author} {\bibfnamefont {A.}~\bibnamefont
  {Zhidenko}},\ }\bibfield  {title} {\enquote {\bibinfo {title} {{First few
  overtones probe the event horizon geometry}},}\ }\href@noop {} {\  (\bibinfo
  {year} {2022})},\ \Eprint {http://arxiv.org/abs/2209.00679} {arXiv:2209.00679
  [gr-qc]} \BibitemShut {NoStop}%
\bibitem [{\citenamefont {Meidam}\ \emph {et~al.}(2014)\citenamefont {Meidam},
  \citenamefont {Agathos}, \citenamefont {Van Den~Broeck}, \citenamefont
  {Veitch},\ and\ \citenamefont {Sathyaprakash}}]{Meidam:2014jpa}%
  \BibitemOpen
  \bibfield  {author} {\bibinfo {author} {\bibfnamefont {J.}~\bibnamefont
  {Meidam}}, \bibinfo {author} {\bibfnamefont {M.}~\bibnamefont {Agathos}},
  \bibinfo {author} {\bibfnamefont {C.}~\bibnamefont {Van Den~Broeck}},
  \bibinfo {author} {\bibfnamefont {J.}~\bibnamefont {Veitch}}, \ and\ \bibinfo
  {author} {\bibfnamefont {B.~S.}\ \bibnamefont {Sathyaprakash}},\ }\bibfield
  {title} {\enquote {\bibinfo {title} {{Testing the no-hair theorem with black
  hole ringdowns using TIGER}},}\ }\href {\doibase 10.1103/PhysRevD.90.064009}
  {\bibfield  {journal} {\bibinfo  {journal} {Phys. Rev. D}\ }\textbf {\bibinfo
  {volume} {90}},\ \bibinfo {pages} {064009} (\bibinfo {year} {2014})},\
  \Eprint {http://arxiv.org/abs/1406.3201} {arXiv:1406.3201 [gr-qc]}
  \BibitemShut {NoStop}%
\bibitem [{\citenamefont {Maselli}\ \emph {et~al.}(2020)\citenamefont
  {Maselli}, \citenamefont {Pani}, \citenamefont {Gualtieri},\ and\
  \citenamefont {Berti}}]{Maselli:2019mjd}%
  \BibitemOpen
  \bibfield  {author} {\bibinfo {author} {\bibfnamefont {Andrea}\ \bibnamefont
  {Maselli}}, \bibinfo {author} {\bibfnamefont {Paolo}\ \bibnamefont {Pani}},
  \bibinfo {author} {\bibfnamefont {Leonardo}\ \bibnamefont {Gualtieri}}, \
  and\ \bibinfo {author} {\bibfnamefont {Emanuele}\ \bibnamefont {Berti}},\
  }\bibfield  {title} {\enquote {\bibinfo {title} {{Parametrized ringdown spin
  expansion coefficients: a data-analysis framework for black-hole spectroscopy
  with multiple events}},}\ }\href {\doibase 10.1103/PhysRevD.101.024043}
  {\bibfield  {journal} {\bibinfo  {journal} {Phys. Rev. D}\ }\textbf {\bibinfo
  {volume} {101}},\ \bibinfo {pages} {024043} (\bibinfo {year} {2020})},\
  \Eprint {http://arxiv.org/abs/1910.12893} {arXiv:1910.12893 [gr-qc]}
  \BibitemShut {NoStop}%
\bibitem [{\citenamefont {Carullo}(2021)}]{Carullo:2021dui}%
  \BibitemOpen
  \bibfield  {author} {\bibinfo {author} {\bibfnamefont {Gregorio}\
  \bibnamefont {Carullo}},\ }\bibfield  {title} {\enquote {\bibinfo {title}
  {{Enhancing modified gravity detection from gravitational-wave observations
  using the parametrized ringdown spin expansion coeffcients formalism}},}\
  }\href {\doibase 10.1103/PhysRevD.103.124043} {\bibfield  {journal} {\bibinfo
   {journal} {Phys. Rev. D}\ }\textbf {\bibinfo {volume} {103}},\ \bibinfo
  {pages} {124043} (\bibinfo {year} {2021})},\ \Eprint
  {http://arxiv.org/abs/2102.05939} {arXiv:2102.05939 [gr-qc]} \BibitemShut
  {NoStop}%
\bibitem [{\citenamefont {Yang}\ \emph {et~al.}(2017)\citenamefont {Yang},
  \citenamefont {Yagi}, \citenamefont {Blackman}, \citenamefont {Lehner},
  \citenamefont {Paschalidis}, \citenamefont {Pretorius},\ and\ \citenamefont
  {Yunes}}]{Yang:2017zxs}%
  \BibitemOpen
  \bibfield  {author} {\bibinfo {author} {\bibfnamefont {Huan}\ \bibnamefont
  {Yang}}, \bibinfo {author} {\bibfnamefont {Kent}\ \bibnamefont {Yagi}},
  \bibinfo {author} {\bibfnamefont {Jonathan}\ \bibnamefont {Blackman}},
  \bibinfo {author} {\bibfnamefont {Luis}\ \bibnamefont {Lehner}}, \bibinfo
  {author} {\bibfnamefont {Vasileios}\ \bibnamefont {Paschalidis}}, \bibinfo
  {author} {\bibfnamefont {Frans}\ \bibnamefont {Pretorius}}, \ and\ \bibinfo
  {author} {\bibfnamefont {Nicol\'as}\ \bibnamefont {Yunes}},\ }\bibfield
  {title} {\enquote {\bibinfo {title} {{Black hole spectroscopy with coherent
  mode stacking}},}\ }\href {\doibase 10.1103/PhysRevLett.118.161101}
  {\bibfield  {journal} {\bibinfo  {journal} {Phys. Rev. Lett.}\ }\textbf
  {\bibinfo {volume} {118}},\ \bibinfo {pages} {161101} (\bibinfo {year}
  {2017})},\ \Eprint {http://arxiv.org/abs/1701.05808} {arXiv:1701.05808
  [gr-qc]} \BibitemShut {NoStop}%
\bibitem [{\citenamefont {Press}(1971)}]{Press:1971wr}%
  \BibitemOpen
  \bibfield  {author} {\bibinfo {author} {\bibfnamefont {William~H.}\
  \bibnamefont {Press}},\ }\bibfield  {title} {\enquote {\bibinfo {title}
  {{Long Wave Trains of Gravitational Waves from a Vibrating Black Hole}},}\
  }\href {\doibase 10.1086/180849} {\bibfield  {journal} {\bibinfo  {journal}
  {Astrophys. J. Lett.}\ }\textbf {\bibinfo {volume} {170}},\ \bibinfo {pages}
  {L105--L108} (\bibinfo {year} {1971})}\BibitemShut {NoStop}%
\bibitem [{\citenamefont {Schutz}\ and\ \citenamefont
  {Will}(1985)}]{Schutz:1985km}%
  \BibitemOpen
  \bibfield  {author} {\bibinfo {author} {\bibfnamefont {Bernard~F.}\
  \bibnamefont {Schutz}}\ and\ \bibinfo {author} {\bibfnamefont {Clifford~M.}\
  \bibnamefont {Will}},\ }\bibfield  {title} {\enquote {\bibinfo {title}
  {{Black Hole Normal Modes: a Semianalytical Approach}},}\ }\href {\doibase
  10.1086/184453} {\bibfield  {journal} {\bibinfo  {journal} {Astrophys. J.
  Lett.}\ }\textbf {\bibinfo {volume} {291}},\ \bibinfo {pages} {L33--L36}
  (\bibinfo {year} {1985})}\BibitemShut {NoStop}%
\bibitem [{\citenamefont {Cardoso}\ \emph {et~al.}(2009)\citenamefont
  {Cardoso}, \citenamefont {Miranda}, \citenamefont {Berti}, \citenamefont
  {Witek},\ and\ \citenamefont {Zanchin}}]{Cardoso:2008bp}%
  \BibitemOpen
  \bibfield  {author} {\bibinfo {author} {\bibfnamefont {Vitor}\ \bibnamefont
  {Cardoso}}, \bibinfo {author} {\bibfnamefont {Alex~S.}\ \bibnamefont
  {Miranda}}, \bibinfo {author} {\bibfnamefont {Emanuele}\ \bibnamefont
  {Berti}}, \bibinfo {author} {\bibfnamefont {Helvi}\ \bibnamefont {Witek}}, \
  and\ \bibinfo {author} {\bibfnamefont {Vilson~T.}\ \bibnamefont {Zanchin}},\
  }\bibfield  {title} {\enquote {\bibinfo {title} {{Geodesic stability,
  Lyapunov exponents and quasinormal modes}},}\ }\href {\doibase
  10.1103/PhysRevD.79.064016} {\bibfield  {journal} {\bibinfo  {journal} {Phys.
  Rev. D}\ }\textbf {\bibinfo {volume} {79}},\ \bibinfo {pages} {064016}
  (\bibinfo {year} {2009})},\ \Eprint {http://arxiv.org/abs/0812.1806}
  {arXiv:0812.1806 [hep-th]} \BibitemShut {NoStop}%
\bibitem [{\citenamefont {Iyer}\ and\ \citenamefont
  {Will}(1987)}]{Iyer:1986np}%
  \BibitemOpen
  \bibfield  {author} {\bibinfo {author} {\bibfnamefont {Sai}\ \bibnamefont
  {Iyer}}\ and\ \bibinfo {author} {\bibfnamefont {Clifford~M.}\ \bibnamefont
  {Will}},\ }\bibfield  {title} {\enquote {\bibinfo {title} {{Black Hole Normal
  Modes: A {WKB} Approach. 1. Foundations and Application of a Higher Order
  {WKB} Analysis of Potential Barrier Scattering}},}\ }\href {\doibase
  10.1103/PhysRevD.35.3621} {\bibfield  {journal} {\bibinfo  {journal} {Phys.
  Rev. D}\ }\textbf {\bibinfo {volume} {35}},\ \bibinfo {pages} {3621}
  (\bibinfo {year} {1987})}\BibitemShut {NoStop}%
\bibitem [{\citenamefont {Konoplya}(2003)}]{Konoplya:2003ii}%
  \BibitemOpen
  \bibfield  {author} {\bibinfo {author} {\bibfnamefont {R.~A.}\ \bibnamefont
  {Konoplya}},\ }\bibfield  {title} {\enquote {\bibinfo {title} {{Quasinormal
  behavior of the d-dimensional Schwarzschild black hole and higher order WKB
  approach}},}\ }\href {\doibase 10.1103/PhysRevD.68.024018} {\bibfield
  {journal} {\bibinfo  {journal} {Phys. Rev. D}\ }\textbf {\bibinfo {volume}
  {68}},\ \bibinfo {pages} {024018} (\bibinfo {year} {2003})},\ \Eprint
  {http://arxiv.org/abs/gr-qc/0303052} {arXiv:gr-qc/0303052} \BibitemShut
  {NoStop}%
\bibitem [{\citenamefont {Matyjasek}\ and\ \citenamefont
  {Opala}(2017)}]{Matyjasek:2017psv}%
  \BibitemOpen
  \bibfield  {author} {\bibinfo {author} {\bibfnamefont {Jerzy}\ \bibnamefont
  {Matyjasek}}\ and\ \bibinfo {author} {\bibfnamefont {Micha\l{}}\ \bibnamefont
  {Opala}},\ }\bibfield  {title} {\enquote {\bibinfo {title} {{Quasinormal
  modes of black holes. The improved semianalytic approach}},}\ }\href
  {\doibase 10.1103/PhysRevD.96.024011} {\bibfield  {journal} {\bibinfo
  {journal} {Phys. Rev. D}\ }\textbf {\bibinfo {volume} {96}},\ \bibinfo
  {pages} {024011} (\bibinfo {year} {2017})},\ \Eprint
  {http://arxiv.org/abs/1704.00361} {arXiv:1704.00361 [gr-qc]} \BibitemShut
  {NoStop}%
\bibitem [{\citenamefont {Glampedakis}\ \emph {et~al.}(2017)\citenamefont
  {Glampedakis}, \citenamefont {Pappas}, \citenamefont {Silva},\ and\
  \citenamefont {Berti}}]{Glampedakis:2017dvb}%
  \BibitemOpen
  \bibfield  {author} {\bibinfo {author} {\bibfnamefont {Kostas}\ \bibnamefont
  {Glampedakis}}, \bibinfo {author} {\bibfnamefont {George}\ \bibnamefont
  {Pappas}}, \bibinfo {author} {\bibfnamefont {Hector~O.}\ \bibnamefont
  {Silva}}, \ and\ \bibinfo {author} {\bibfnamefont {Emanuele}\ \bibnamefont
  {Berti}},\ }\bibfield  {title} {\enquote {\bibinfo {title} {{Post-Kerr black
  hole spectroscopy}},}\ }\href {\doibase 10.1103/PhysRevD.96.064054}
  {\bibfield  {journal} {\bibinfo  {journal} {Phys. Rev. D}\ }\textbf {\bibinfo
  {volume} {96}},\ \bibinfo {pages} {064054} (\bibinfo {year} {2017})},\
  \Eprint {http://arxiv.org/abs/1706.07658} {arXiv:1706.07658 [gr-qc]}
  \BibitemShut {NoStop}%
\bibitem [{\citenamefont {Glampedakis}\ and\ \citenamefont
  {Silva}(2019)}]{Glampedakis:2019dqh}%
  \BibitemOpen
  \bibfield  {author} {\bibinfo {author} {\bibfnamefont {Kostas}\ \bibnamefont
  {Glampedakis}}\ and\ \bibinfo {author} {\bibfnamefont {Hector~O.}\
  \bibnamefont {Silva}},\ }\bibfield  {title} {\enquote {\bibinfo {title}
  {{Eikonal quasinormal modes of black holes beyond General Relativity}},}\
  }\href {\doibase 10.1103/PhysRevD.100.044040} {\bibfield  {journal} {\bibinfo
   {journal} {Phys. Rev. D}\ }\textbf {\bibinfo {volume} {100}},\ \bibinfo
  {pages} {044040} (\bibinfo {year} {2019})},\ \Eprint
  {http://arxiv.org/abs/1906.05455} {arXiv:1906.05455 [gr-qc]} \BibitemShut
  {NoStop}%
\bibitem [{\citenamefont {Silva}\ and\ \citenamefont
  {Glampedakis}(2020)}]{Silva:2019scu}%
  \BibitemOpen
  \bibfield  {author} {\bibinfo {author} {\bibfnamefont {Hector~O.}\
  \bibnamefont {Silva}}\ and\ \bibinfo {author} {\bibfnamefont {Kostas}\
  \bibnamefont {Glampedakis}},\ }\bibfield  {title} {\enquote {\bibinfo {title}
  {{Eikonal quasinormal modes of black holes beyond general relativity. II.
  Generalized scalar-tensor perturbations}},}\ }\href {\doibase
  10.1103/PhysRevD.101.044051} {\bibfield  {journal} {\bibinfo  {journal}
  {Phys. Rev. D}\ }\textbf {\bibinfo {volume} {101}},\ \bibinfo {pages}
  {044051} (\bibinfo {year} {2020})},\ \Eprint
  {http://arxiv.org/abs/1912.09286} {arXiv:1912.09286 [gr-qc]} \BibitemShut
  {NoStop}%
\bibitem [{\citenamefont {Bryant}\ \emph {et~al.}(2021)\citenamefont {Bryant},
  \citenamefont {Silva}, \citenamefont {Yagi},\ and\ \citenamefont
  {Glampedakis}}]{Bryant:2021xdh}%
  \BibitemOpen
  \bibfield  {author} {\bibinfo {author} {\bibfnamefont {Albert}\ \bibnamefont
  {Bryant}}, \bibinfo {author} {\bibfnamefont {Hector~O.}\ \bibnamefont
  {Silva}}, \bibinfo {author} {\bibfnamefont {Kent}\ \bibnamefont {Yagi}}, \
  and\ \bibinfo {author} {\bibfnamefont {Kostas}\ \bibnamefont {Glampedakis}},\
  }\bibfield  {title} {\enquote {\bibinfo {title} {{Eikonal quasinormal modes
  of black holes beyond general relativity. III. Scalar Gauss-Bonnet
  gravity}},}\ }\href {\doibase 10.1103/PhysRevD.104.044051} {\bibfield
  {journal} {\bibinfo  {journal} {Phys. Rev. D}\ }\textbf {\bibinfo {volume}
  {104}},\ \bibinfo {pages} {044051} (\bibinfo {year} {2021})},\ \Eprint
  {http://arxiv.org/abs/2106.09657} {arXiv:2106.09657 [gr-qc]} \BibitemShut
  {NoStop}%
\bibitem [{\citenamefont {Akiyama}\ \emph
  {et~al.}(2019{\natexlab{a}})\citenamefont {Akiyama} \emph
  {et~al.}}]{EventHorizonTelescope:2019dse}%
  \BibitemOpen
  \bibfield  {author} {\bibinfo {author} {\bibfnamefont {Kazunori}\
  \bibnamefont {Akiyama}} \emph {et~al.} (\bibinfo {collaboration} {Event
  Horizon Telescope}),\ }\bibfield  {title} {\enquote {\bibinfo {title} {{First
  M87 Event Horizon Telescope Results. I. The Shadow of the Supermassive Black
  Hole}},}\ }\href {\doibase 10.3847/2041-8213/ab0ec7} {\bibfield  {journal}
  {\bibinfo  {journal} {Astrophys. J. Lett.}\ }\textbf {\bibinfo {volume}
  {875}},\ \bibinfo {pages} {L1} (\bibinfo {year} {2019}{\natexlab{a}})},\
  \Eprint {http://arxiv.org/abs/1906.11238} {arXiv:1906.11238 [astro-ph.GA]}
  \BibitemShut {NoStop}%
\bibitem [{\citenamefont {Akiyama}\ \emph
  {et~al.}(2019{\natexlab{b}})\citenamefont {Akiyama} \emph
  {et~al.}}]{EventHorizonTelescope:2019ggy}%
  \BibitemOpen
  \bibfield  {author} {\bibinfo {author} {\bibfnamefont {Kazunori}\
  \bibnamefont {Akiyama}} \emph {et~al.} (\bibinfo {collaboration} {Event
  Horizon Telescope}),\ }\bibfield  {title} {\enquote {\bibinfo {title} {{First
  M87 Event Horizon Telescope Results. VI. The Shadow and Mass of the Central
  Black Hole}},}\ }\href {\doibase 10.3847/2041-8213/ab1141} {\bibfield
  {journal} {\bibinfo  {journal} {Astrophys. J. Lett.}\ }\textbf {\bibinfo
  {volume} {875}},\ \bibinfo {pages} {L6} (\bibinfo {year}
  {2019}{\natexlab{b}})},\ \Eprint {http://arxiv.org/abs/1906.11243}
  {arXiv:1906.11243 [astro-ph.GA]} \BibitemShut {NoStop}%
\bibitem [{\citenamefont {V\"olkel}\ \emph {et~al.}(2021)\citenamefont
  {V\"olkel}, \citenamefont {Barausse}, \citenamefont {Franchini},\ and\
  \citenamefont {Broderick}}]{Volkel:2020xlc}%
  \BibitemOpen
  \bibfield  {author} {\bibinfo {author} {\bibfnamefont {Sebastian~H.}\
  \bibnamefont {V\"olkel}}, \bibinfo {author} {\bibfnamefont {Enrico}\
  \bibnamefont {Barausse}}, \bibinfo {author} {\bibfnamefont {Nicola}\
  \bibnamefont {Franchini}}, \ and\ \bibinfo {author} {\bibfnamefont
  {Avery~E.}\ \bibnamefont {Broderick}},\ }\bibfield  {title} {\enquote
  {\bibinfo {title} {{EHT tests of the strong-field regime of general
  relativity}},}\ }\href {\doibase 10.1088/1361-6382/ac27ed} {\bibfield
  {journal} {\bibinfo  {journal} {Class. Quant. Grav.}\ }\textbf {\bibinfo
  {volume} {38}},\ \bibinfo {pages} {21LT01} (\bibinfo {year} {2021})},\
  \Eprint {http://arxiv.org/abs/2011.06812} {arXiv:2011.06812 [gr-qc]}
  \BibitemShut {NoStop}%
\bibitem [{\citenamefont {Kocherlakota}\ \emph {et~al.}(2021)\citenamefont
  {Kocherlakota} \emph {et~al.}}]{EventHorizonTelescope:2021dqv}%
  \BibitemOpen
  \bibfield  {author} {\bibinfo {author} {\bibfnamefont {Prashant}\
  \bibnamefont {Kocherlakota}} \emph {et~al.} (\bibinfo {collaboration} {Event
  Horizon Telescope}),\ }\bibfield  {title} {\enquote {\bibinfo {title}
  {{Constraints on black-hole charges with the 2017 EHT observations of
  M87*}},}\ }\href {\doibase 10.1103/PhysRevD.103.104047} {\bibfield  {journal}
  {\bibinfo  {journal} {Phys. Rev. D}\ }\textbf {\bibinfo {volume} {103}},\
  \bibinfo {pages} {104047} (\bibinfo {year} {2021})},\ \Eprint
  {http://arxiv.org/abs/2105.09343} {arXiv:2105.09343 [gr-qc]} \BibitemShut
  {NoStop}%
\bibitem [{\citenamefont {Regge}\ and\ \citenamefont
  {Wheeler}(1957)}]{Regge:1957td}%
  \BibitemOpen
  \bibfield  {author} {\bibinfo {author} {\bibfnamefont {Tullio}\ \bibnamefont
  {Regge}}\ and\ \bibinfo {author} {\bibfnamefont {John~A.}\ \bibnamefont
  {Wheeler}},\ }\bibfield  {title} {\enquote {\bibinfo {title} {{Stability of a
  Schwarzschild singularity}},}\ }\href {\doibase 10.1103/PhysRev.108.1063}
  {\bibfield  {journal} {\bibinfo  {journal} {Phys. Rev.}\ }\textbf {\bibinfo
  {volume} {108}},\ \bibinfo {pages} {1063--1069} (\bibinfo {year}
  {1957})}\BibitemShut {NoStop}%
\bibitem [{\citenamefont {Zerilli}(1970)}]{Zerilli:1970se}%
  \BibitemOpen
  \bibfield  {author} {\bibinfo {author} {\bibfnamefont {Frank~J.}\
  \bibnamefont {Zerilli}},\ }\bibfield  {title} {\enquote {\bibinfo {title}
  {{Effective potential for even parity Regge-Wheeler gravitational
  perturbation equations}},}\ }\href {\doibase 10.1103/PhysRevLett.24.737}
  {\bibfield  {journal} {\bibinfo  {journal} {Phys. Rev. Lett.}\ }\textbf
  {\bibinfo {volume} {24}},\ \bibinfo {pages} {737--738} (\bibinfo {year}
  {1970})}\BibitemShut {NoStop}%
\bibitem [{\citenamefont {Teukolsky}(1973)}]{Teukolsky:1973ha}%
  \BibitemOpen
  \bibfield  {author} {\bibinfo {author} {\bibfnamefont {Saul~A.}\ \bibnamefont
  {Teukolsky}},\ }\bibfield  {title} {\enquote {\bibinfo {title}
  {{Perturbations of a rotating black hole. 1. Fundamental equations for
  gravitational electromagnetic and neutrino field perturbations}},}\ }\href
  {\doibase 10.1086/152444} {\bibfield  {journal} {\bibinfo  {journal}
  {Astrophys. J.}\ }\textbf {\bibinfo {volume} {185}},\ \bibinfo {pages}
  {635--647} (\bibinfo {year} {1973})}\BibitemShut {NoStop}%
\bibitem [{\citenamefont {Kokkotas}\ and\ \citenamefont
  {Schmidt}(1999)}]{Kokkotas:1999bd}%
  \BibitemOpen
  \bibfield  {author} {\bibinfo {author} {\bibfnamefont {Kostas~D.}\
  \bibnamefont {Kokkotas}}\ and\ \bibinfo {author} {\bibfnamefont {Bernd~G.}\
  \bibnamefont {Schmidt}},\ }\bibfield  {title} {\enquote {\bibinfo {title}
  {{Quasinormal modes of stars and black holes}},}\ }\href {\doibase
  10.12942/lrr-1999-2} {\bibfield  {journal} {\bibinfo  {journal} {Living Rev.
  Rel.}\ }\textbf {\bibinfo {volume} {2}},\ \bibinfo {pages} {2} (\bibinfo
  {year} {1999})},\ \Eprint {http://arxiv.org/abs/gr-qc/9909058}
  {arXiv:gr-qc/9909058} \BibitemShut {NoStop}%
\bibitem [{\citenamefont {Nollert}(1999)}]{Nollert:1999ji}%
  \BibitemOpen
  \bibfield  {author} {\bibinfo {author} {\bibfnamefont {Hans-Peter}\
  \bibnamefont {Nollert}},\ }\bibfield  {title} {\enquote {\bibinfo {title}
  {{TOPICAL REVIEW: Quasinormal modes: the characteristic `sound' of black
  holes and neutron stars}},}\ }\href {\doibase 10.1088/0264-9381/16/12/201}
  {\bibfield  {journal} {\bibinfo  {journal} {Class. Quant. Grav.}\ }\textbf
  {\bibinfo {volume} {16}},\ \bibinfo {pages} {R159--R216} (\bibinfo {year}
  {1999})}\BibitemShut {NoStop}%
\bibitem [{\citenamefont {Berti}\ \emph {et~al.}(2009)\citenamefont {Berti},
  \citenamefont {Cardoso},\ and\ \citenamefont {Starinets}}]{Berti:2009kk}%
  \BibitemOpen
  \bibfield  {author} {\bibinfo {author} {\bibfnamefont {Emanuele}\
  \bibnamefont {Berti}}, \bibinfo {author} {\bibfnamefont {Vitor}\ \bibnamefont
  {Cardoso}}, \ and\ \bibinfo {author} {\bibfnamefont {Andrei~O.}\ \bibnamefont
  {Starinets}},\ }\bibfield  {title} {\enquote {\bibinfo {title} {{Quasinormal
  modes of black holes and black branes}},}\ }\href {\doibase
  10.1088/0264-9381/26/16/163001} {\bibfield  {journal} {\bibinfo  {journal}
  {Class. Quant. Grav.}\ }\textbf {\bibinfo {volume} {26}},\ \bibinfo {pages}
  {163001} (\bibinfo {year} {2009})},\ \Eprint {http://arxiv.org/abs/0905.2975}
  {arXiv:0905.2975 [gr-qc]} \BibitemShut {NoStop}%
\bibitem [{\citenamefont {Konoplya}\ and\ \citenamefont
  {Zhidenko}(2011)}]{Konoplya:2011qq}%
  \BibitemOpen
  \bibfield  {author} {\bibinfo {author} {\bibfnamefont {R.~A.}\ \bibnamefont
  {Konoplya}}\ and\ \bibinfo {author} {\bibfnamefont {A.}~\bibnamefont
  {Zhidenko}},\ }\bibfield  {title} {\enquote {\bibinfo {title} {{Quasinormal
  modes of black holes: From astrophysics to string theory}},}\ }\href
  {\doibase 10.1103/RevModPhys.83.793} {\bibfield  {journal} {\bibinfo
  {journal} {Rev. Mod. Phys.}\ }\textbf {\bibinfo {volume} {83}},\ \bibinfo
  {pages} {793--836} (\bibinfo {year} {2011})},\ \Eprint
  {http://arxiv.org/abs/1102.4014} {arXiv:1102.4014 [gr-qc]} \BibitemShut
  {NoStop}%
\bibitem [{\citenamefont {Pani}(2013)}]{Pani:2013pma}%
  \BibitemOpen
  \bibfield  {author} {\bibinfo {author} {\bibfnamefont {Paolo}\ \bibnamefont
  {Pani}},\ }\bibfield  {title} {\enquote {\bibinfo {title} {{Advanced Methods
  in Black-Hole Perturbation Theory}},}\ }\href {\doibase
  10.1142/S0217751X13400186} {\bibfield  {journal} {\bibinfo  {journal} {Int.
  J. Mod. Phys. A}\ }\textbf {\bibinfo {volume} {28}},\ \bibinfo {pages}
  {1340018} (\bibinfo {year} {2013})},\ \Eprint
  {http://arxiv.org/abs/1305.6759} {arXiv:1305.6759 [gr-qc]} \BibitemShut
  {NoStop}%
\bibitem [{\citenamefont {Cardoso}\ and\ \citenamefont
  {Gualtieri}(2009)}]{Cardoso:2009pk}%
  \BibitemOpen
  \bibfield  {author} {\bibinfo {author} {\bibfnamefont {Vitor}\ \bibnamefont
  {Cardoso}}\ and\ \bibinfo {author} {\bibfnamefont {Leonardo}\ \bibnamefont
  {Gualtieri}},\ }\bibfield  {title} {\enquote {\bibinfo {title}
  {{Perturbations of Schwarzschild black holes in Dynamical Chern-Simons
  modified gravity}},}\ }\href {\doibase 10.1103/PhysRevD.81.089903} {\bibfield
   {journal} {\bibinfo  {journal} {Phys. Rev. D}\ }\textbf {\bibinfo {volume}
  {80}},\ \bibinfo {pages} {064008} (\bibinfo {year} {2009})},\ \bibinfo {note}
  {[Erratum: Phys.Rev.D 81, 089903 (2010)]},\ \Eprint
  {http://arxiv.org/abs/0907.5008} {arXiv:0907.5008 [gr-qc]} \BibitemShut
  {NoStop}%
\bibitem [{\citenamefont {Molina}\ \emph {et~al.}(2010)\citenamefont {Molina},
  \citenamefont {Pani}, \citenamefont {Cardoso},\ and\ \citenamefont
  {Gualtieri}}]{Molina:2010fb}%
  \BibitemOpen
  \bibfield  {author} {\bibinfo {author} {\bibfnamefont {C.}~\bibnamefont
  {Molina}}, \bibinfo {author} {\bibfnamefont {Paolo}\ \bibnamefont {Pani}},
  \bibinfo {author} {\bibfnamefont {Vitor}\ \bibnamefont {Cardoso}}, \ and\
  \bibinfo {author} {\bibfnamefont {Leonardo}\ \bibnamefont {Gualtieri}},\
  }\bibfield  {title} {\enquote {\bibinfo {title} {{Gravitational signature of
  Schwarzschild black holes in dynamical Chern-Simons gravity}},}\ }\href
  {\doibase 10.1103/PhysRevD.81.124021} {\bibfield  {journal} {\bibinfo
  {journal} {Phys. Rev. D}\ }\textbf {\bibinfo {volume} {81}},\ \bibinfo
  {pages} {124021} (\bibinfo {year} {2010})},\ \Eprint
  {http://arxiv.org/abs/1004.4007} {arXiv:1004.4007 [gr-qc]} \BibitemShut
  {NoStop}%
\bibitem [{\citenamefont {Bl\'azquez-Salcedo}\ \emph
  {et~al.}(2016)\citenamefont {Bl\'azquez-Salcedo}, \citenamefont {Macedo},
  \citenamefont {Cardoso}, \citenamefont {Ferrari}, \citenamefont {Gualtieri},
  \citenamefont {Khoo}, \citenamefont {Kunz},\ and\ \citenamefont
  {Pani}}]{Blazquez-Salcedo:2016enn}%
  \BibitemOpen
  \bibfield  {author} {\bibinfo {author} {\bibfnamefont {Jose~Luis}\
  \bibnamefont {Bl\'azquez-Salcedo}}, \bibinfo {author} {\bibfnamefont {Caio
  F.~B.}\ \bibnamefont {Macedo}}, \bibinfo {author} {\bibfnamefont {Vitor}\
  \bibnamefont {Cardoso}}, \bibinfo {author} {\bibfnamefont {Valeria}\
  \bibnamefont {Ferrari}}, \bibinfo {author} {\bibfnamefont {Leonardo}\
  \bibnamefont {Gualtieri}}, \bibinfo {author} {\bibfnamefont {Fech~Scen}\
  \bibnamefont {Khoo}}, \bibinfo {author} {\bibfnamefont {Jutta}\ \bibnamefont
  {Kunz}}, \ and\ \bibinfo {author} {\bibfnamefont {Paolo}\ \bibnamefont
  {Pani}},\ }\bibfield  {title} {\enquote {\bibinfo {title} {{Perturbed black
  holes in Einstein-dilaton-Gauss-Bonnet gravity: Stability, ringdown, and
  gravitational-wave emission}},}\ }\href {\doibase 10.1103/PhysRevD.94.104024}
  {\bibfield  {journal} {\bibinfo  {journal} {Phys. Rev. D}\ }\textbf {\bibinfo
  {volume} {94}},\ \bibinfo {pages} {104024} (\bibinfo {year} {2016})},\
  \Eprint {http://arxiv.org/abs/1609.01286} {arXiv:1609.01286 [gr-qc]}
  \BibitemShut {NoStop}%
\bibitem [{\citenamefont {Bl\'azquez-Salcedo}\ \emph
  {et~al.}(2017)\citenamefont {Bl\'azquez-Salcedo}, \citenamefont {Khoo},\ and\
  \citenamefont {Kunz}}]{Blazquez-Salcedo:2017txk}%
  \BibitemOpen
  \bibfield  {author} {\bibinfo {author} {\bibfnamefont {Jose~Luis}\
  \bibnamefont {Bl\'azquez-Salcedo}}, \bibinfo {author} {\bibfnamefont
  {Fech~Scen}\ \bibnamefont {Khoo}}, \ and\ \bibinfo {author} {\bibfnamefont
  {Jutta}\ \bibnamefont {Kunz}},\ }\bibfield  {title} {\enquote {\bibinfo
  {title} {{Quasinormal modes of Einstein-Gauss-Bonnet-dilaton black holes}},}\
  }\href {\doibase 10.1103/PhysRevD.96.064008} {\bibfield  {journal} {\bibinfo
  {journal} {Phys. Rev. D}\ }\textbf {\bibinfo {volume} {96}},\ \bibinfo
  {pages} {064008} (\bibinfo {year} {2017})},\ \Eprint
  {http://arxiv.org/abs/1706.03262} {arXiv:1706.03262 [gr-qc]} \BibitemShut
  {NoStop}%
\bibitem [{\citenamefont {Bl\'azquez-Salcedo}\ \emph
  {et~al.}(2020{\natexlab{a}})\citenamefont {Bl\'azquez-Salcedo}, \citenamefont
  {Doneva}, \citenamefont {Kahlen}, \citenamefont {Kunz}, \citenamefont
  {Nedkova},\ and\ \citenamefont {Yazadjiev}}]{Blazquez-Salcedo:2020rhf}%
  \BibitemOpen
  \bibfield  {author} {\bibinfo {author} {\bibfnamefont {Jose~Luis}\
  \bibnamefont {Bl\'azquez-Salcedo}}, \bibinfo {author} {\bibfnamefont
  {Daniela~D.}\ \bibnamefont {Doneva}}, \bibinfo {author} {\bibfnamefont
  {Sarah}\ \bibnamefont {Kahlen}}, \bibinfo {author} {\bibfnamefont {Jutta}\
  \bibnamefont {Kunz}}, \bibinfo {author} {\bibfnamefont {Petya}\ \bibnamefont
  {Nedkova}}, \ and\ \bibinfo {author} {\bibfnamefont {Stoytcho~S.}\
  \bibnamefont {Yazadjiev}},\ }\bibfield  {title} {\enquote {\bibinfo {title}
  {{Axial perturbations of the scalarized Einstein-Gauss-Bonnet black
  holes}},}\ }\href {\doibase 10.1103/PhysRevD.101.104006} {\bibfield
  {journal} {\bibinfo  {journal} {Phys. Rev. D}\ }\textbf {\bibinfo {volume}
  {101}},\ \bibinfo {pages} {104006} (\bibinfo {year} {2020}{\natexlab{a}})},\
  \Eprint {http://arxiv.org/abs/2003.02862} {arXiv:2003.02862 [gr-qc]}
  \BibitemShut {NoStop}%
\bibitem [{\citenamefont {Bl\'azquez-Salcedo}\ \emph
  {et~al.}(2020{\natexlab{b}})\citenamefont {Bl\'azquez-Salcedo}, \citenamefont
  {Doneva}, \citenamefont {Kahlen}, \citenamefont {Kunz}, \citenamefont
  {Nedkova},\ and\ \citenamefont {Yazadjiev}}]{Blazquez-Salcedo:2020caw}%
  \BibitemOpen
  \bibfield  {author} {\bibinfo {author} {\bibfnamefont {Jose~Luis}\
  \bibnamefont {Bl\'azquez-Salcedo}}, \bibinfo {author} {\bibfnamefont
  {Daniela~D.}\ \bibnamefont {Doneva}}, \bibinfo {author} {\bibfnamefont
  {Sarah}\ \bibnamefont {Kahlen}}, \bibinfo {author} {\bibfnamefont {Jutta}\
  \bibnamefont {Kunz}}, \bibinfo {author} {\bibfnamefont {Petya}\ \bibnamefont
  {Nedkova}}, \ and\ \bibinfo {author} {\bibfnamefont {Stoytcho~S.}\
  \bibnamefont {Yazadjiev}},\ }\bibfield  {title} {\enquote {\bibinfo {title}
  {{Polar quasinormal modes of the scalarized Einstein-Gauss-Bonnet black
  holes}},}\ }\href {\doibase 10.1103/PhysRevD.102.024086} {\bibfield
  {journal} {\bibinfo  {journal} {Phys. Rev. D}\ }\textbf {\bibinfo {volume}
  {102}},\ \bibinfo {pages} {024086} (\bibinfo {year} {2020}{\natexlab{b}})},\
  \Eprint {http://arxiv.org/abs/2006.06006} {arXiv:2006.06006 [gr-qc]}
  \BibitemShut {NoStop}%
\bibitem [{\citenamefont {Franchini}\ \emph {et~al.}(2021)\citenamefont
  {Franchini}, \citenamefont {Herrero-Valea},\ and\ \citenamefont
  {Barausse}}]{Franchini:2021bpt}%
  \BibitemOpen
  \bibfield  {author} {\bibinfo {author} {\bibfnamefont {Nicola}\ \bibnamefont
  {Franchini}}, \bibinfo {author} {\bibfnamefont {Mario}\ \bibnamefont
  {Herrero-Valea}}, \ and\ \bibinfo {author} {\bibfnamefont {Enrico}\
  \bibnamefont {Barausse}},\ }\bibfield  {title} {\enquote {\bibinfo {title}
  {{Relation between general relativity and a class of Ho\v{r}ava gravity
  theories}},}\ }\href {\doibase 10.1103/PhysRevD.103.084012} {\bibfield
  {journal} {\bibinfo  {journal} {Phys. Rev. D}\ }\textbf {\bibinfo {volume}
  {103}},\ \bibinfo {pages} {084012} (\bibinfo {year} {2021})},\ \Eprint
  {http://arxiv.org/abs/2103.00929} {arXiv:2103.00929 [gr-qc]} \BibitemShut
  {NoStop}%
\bibitem [{\citenamefont {Wagle}\ \emph {et~al.}(2022)\citenamefont {Wagle},
  \citenamefont {Yunes},\ and\ \citenamefont {Silva}}]{Wagle:2021tam}%
  \BibitemOpen
  \bibfield  {author} {\bibinfo {author} {\bibfnamefont {Pratik}\ \bibnamefont
  {Wagle}}, \bibinfo {author} {\bibfnamefont {Nicolas}\ \bibnamefont {Yunes}},
  \ and\ \bibinfo {author} {\bibfnamefont {Hector~O.}\ \bibnamefont {Silva}},\
  }\bibfield  {title} {\enquote {\bibinfo {title} {{Quasinormal modes of
  slowly-rotating black holes in dynamical Chern-Simons gravity}},}\ }\href
  {\doibase 10.1103/PhysRevD.105.124003} {\bibfield  {journal} {\bibinfo
  {journal} {Phys. Rev. D}\ }\textbf {\bibinfo {volume} {105}},\ \bibinfo
  {pages} {124003} (\bibinfo {year} {2022})},\ \Eprint
  {http://arxiv.org/abs/2103.09913} {arXiv:2103.09913 [gr-qc]} \BibitemShut
  {NoStop}%
\bibitem [{\citenamefont {Srivastava}\ \emph {et~al.}(2021)\citenamefont
  {Srivastava}, \citenamefont {Chen},\ and\ \citenamefont
  {Shankaranarayanan}}]{Srivastava:2021imr}%
  \BibitemOpen
  \bibfield  {author} {\bibinfo {author} {\bibfnamefont {Manu}\ \bibnamefont
  {Srivastava}}, \bibinfo {author} {\bibfnamefont {Yanbei}\ \bibnamefont
  {Chen}}, \ and\ \bibinfo {author} {\bibfnamefont {S.}~\bibnamefont
  {Shankaranarayanan}},\ }\bibfield  {title} {\enquote {\bibinfo {title}
  {{Analytical computation of quasinormal modes of slowly rotating black holes
  in dynamical Chern-Simons gravity}},}\ }\href {\doibase
  10.1103/PhysRevD.104.064034} {\bibfield  {journal} {\bibinfo  {journal}
  {Phys. Rev. D}\ }\textbf {\bibinfo {volume} {104}},\ \bibinfo {pages}
  {064034} (\bibinfo {year} {2021})},\ \Eprint
  {http://arxiv.org/abs/2106.06209} {arXiv:2106.06209 [gr-qc]} \BibitemShut
  {NoStop}%
\bibitem [{\citenamefont {Pierini}\ and\ \citenamefont
  {Gualtieri}(2021)}]{Pierini:2021jxd}%
  \BibitemOpen
  \bibfield  {author} {\bibinfo {author} {\bibfnamefont {Lorenzo}\ \bibnamefont
  {Pierini}}\ and\ \bibinfo {author} {\bibfnamefont {Leonardo}\ \bibnamefont
  {Gualtieri}},\ }\bibfield  {title} {\enquote {\bibinfo {title} {{Quasi-normal
  modes of rotating black holes in Einstein-dilaton Gauss-Bonnet gravity: the
  first order in rotation}},}\ }\href {\doibase 10.1103/PhysRevD.103.124017}
  {\bibfield  {journal} {\bibinfo  {journal} {Phys. Rev. D}\ }\textbf {\bibinfo
  {volume} {103}},\ \bibinfo {pages} {124017} (\bibinfo {year} {2021})},\
  \Eprint {http://arxiv.org/abs/2103.09870} {arXiv:2103.09870 [gr-qc]}
  \BibitemShut {NoStop}%
\bibitem [{\citenamefont {Pierini}\ and\ \citenamefont
  {Gualtieri}(2022)}]{Pierini:2022eim}%
  \BibitemOpen
  \bibfield  {author} {\bibinfo {author} {\bibfnamefont {Lorenzo}\ \bibnamefont
  {Pierini}}\ and\ \bibinfo {author} {\bibfnamefont {Leonardo}\ \bibnamefont
  {Gualtieri}},\ }\bibfield  {title} {\enquote {\bibinfo {title} {{Quasi-normal
  modes of rotating black holes in Einstein-dilaton Gauss-Bonnet gravity: the
  second order in rotation}},}\ }\href@noop {} {\  (\bibinfo {year} {2022})},\
  \Eprint {http://arxiv.org/abs/2207.11267} {arXiv:2207.11267 [gr-qc]}
  \BibitemShut {NoStop}%
\bibitem [{\citenamefont {Cano}\ \emph {et~al.}(2020)\citenamefont {Cano},
  \citenamefont {Fransen},\ and\ \citenamefont {Hertog}}]{Cano:2020cao}%
  \BibitemOpen
  \bibfield  {author} {\bibinfo {author} {\bibfnamefont {Pablo~A.}\
  \bibnamefont {Cano}}, \bibinfo {author} {\bibfnamefont {Kwinten}\
  \bibnamefont {Fransen}}, \ and\ \bibinfo {author} {\bibfnamefont {Thomas}\
  \bibnamefont {Hertog}},\ }\bibfield  {title} {\enquote {\bibinfo {title}
  {{Ringing of rotating black holes in higher-derivative gravity}},}\ }\href
  {\doibase 10.1103/PhysRevD.102.044047} {\bibfield  {journal} {\bibinfo
  {journal} {Phys. Rev. D}\ }\textbf {\bibinfo {volume} {102}},\ \bibinfo
  {pages} {044047} (\bibinfo {year} {2020})},\ \Eprint
  {http://arxiv.org/abs/2005.03671} {arXiv:2005.03671 [gr-qc]} \BibitemShut
  {NoStop}%
\bibitem [{\citenamefont {Cano}\ \emph {et~al.}(2022)\citenamefont {Cano},
  \citenamefont {Fransen}, \citenamefont {Hertog},\ and\ \citenamefont
  {Maenaut}}]{Cano:2021myl}%
  \BibitemOpen
  \bibfield  {author} {\bibinfo {author} {\bibfnamefont {Pablo~A.}\
  \bibnamefont {Cano}}, \bibinfo {author} {\bibfnamefont {Kwinten}\
  \bibnamefont {Fransen}}, \bibinfo {author} {\bibfnamefont {Thomas}\
  \bibnamefont {Hertog}}, \ and\ \bibinfo {author} {\bibfnamefont {Simon}\
  \bibnamefont {Maenaut}},\ }\bibfield  {title} {\enquote {\bibinfo {title}
  {{Gravitational ringing of rotating black holes in higher-derivative
  gravity}},}\ }\href {\doibase 10.1103/PhysRevD.105.024064} {\bibfield
  {journal} {\bibinfo  {journal} {Phys. Rev. D}\ }\textbf {\bibinfo {volume}
  {105}},\ \bibinfo {pages} {024064} (\bibinfo {year} {2022})},\ \Eprint
  {http://arxiv.org/abs/2110.11378} {arXiv:2110.11378 [gr-qc]} \BibitemShut
  {NoStop}%
\bibitem [{\citenamefont {Li}\ \emph {et~al.}(2022)\citenamefont {Li},
  \citenamefont {Wagle}, \citenamefont {Chen},\ and\ \citenamefont
  {Yunes}}]{Li:2022pcy}%
  \BibitemOpen
  \bibfield  {author} {\bibinfo {author} {\bibfnamefont {Dongjun}\ \bibnamefont
  {Li}}, \bibinfo {author} {\bibfnamefont {Pratik}\ \bibnamefont {Wagle}},
  \bibinfo {author} {\bibfnamefont {Yanbei}\ \bibnamefont {Chen}}, \ and\
  \bibinfo {author} {\bibfnamefont {Nicol\'as}\ \bibnamefont {Yunes}},\
  }\bibfield  {title} {\enquote {\bibinfo {title} {{Perturbations of spinning
  black holes beyond General Relativity: Modified Teukolsky equation}},}\
  }\href@noop {} {\  (\bibinfo {year} {2022})},\ \Eprint
  {http://arxiv.org/abs/2206.10652} {arXiv:2206.10652 [gr-qc]} \BibitemShut
  {NoStop}%
\bibitem [{\citenamefont {Hussain}\ and\ \citenamefont
  {Zimmerman}(2022)}]{Hussain:2022ins}%
  \BibitemOpen
  \bibfield  {author} {\bibinfo {author} {\bibfnamefont {Asad}\ \bibnamefont
  {Hussain}}\ and\ \bibinfo {author} {\bibfnamefont {Aaron}\ \bibnamefont
  {Zimmerman}},\ }\bibfield  {title} {\enquote {\bibinfo {title} {{An approach
  to computing spectral shifts for black holes beyond Kerr}},}\ }\href@noop {}
  {\  (\bibinfo {year} {2022})},\ \Eprint {http://arxiv.org/abs/2206.10653}
  {arXiv:2206.10653 [gr-qc]} \BibitemShut {NoStop}%
\bibitem [{\citenamefont {Kimura}(2020)}]{Kimura:2020mrh}%
  \BibitemOpen
  \bibfield  {author} {\bibinfo {author} {\bibfnamefont {Masashi}\ \bibnamefont
  {Kimura}},\ }\bibfield  {title} {\enquote {\bibinfo {title} {{Note on the
  parametrized black hole quasinormal ringdown formalism}},}\ }\href {\doibase
  10.1103/PhysRevD.101.064031} {\bibfield  {journal} {\bibinfo  {journal}
  {Phys. Rev. D}\ }\textbf {\bibinfo {volume} {101}},\ \bibinfo {pages}
  {064031} (\bibinfo {year} {2020})},\ \Eprint
  {http://arxiv.org/abs/2001.09613} {arXiv:2001.09613 [gr-qc]} \BibitemShut
  {NoStop}%
\bibitem [{\citenamefont {Konoplya}\ \emph {et~al.}(2019)\citenamefont
  {Konoplya}, \citenamefont {Zhidenko},\ and\ \citenamefont
  {Zinhailo}}]{Konoplya:2019hlu}%
  \BibitemOpen
  \bibfield  {author} {\bibinfo {author} {\bibfnamefont {R.~A.}\ \bibnamefont
  {Konoplya}}, \bibinfo {author} {\bibfnamefont {A.}~\bibnamefont {Zhidenko}},
  \ and\ \bibinfo {author} {\bibfnamefont {A.~F.}\ \bibnamefont {Zinhailo}},\
  }\bibfield  {title} {\enquote {\bibinfo {title} {{Higher order WKB formula
  for quasinormal modes and grey-body factors: recipes for quick and accurate
  calculations}},}\ }\href {\doibase 10.1088/1361-6382/ab2e25} {\bibfield
  {journal} {\bibinfo  {journal} {Class. Quant. Grav.}\ }\textbf {\bibinfo
  {volume} {36}},\ \bibinfo {pages} {155002} (\bibinfo {year} {2019})},\
  \Eprint {http://arxiv.org/abs/1904.10333} {arXiv:1904.10333 [gr-qc]}
  \BibitemShut {NoStop}%
\bibitem [{\citenamefont {Foreman-Mackey}\ \emph {et~al.}(2013)\citenamefont
  {Foreman-Mackey}, \citenamefont {Hogg}, \citenamefont {Lang},\ and\
  \citenamefont {Goodman}}]{Foreman-Mackey:2012any}%
  \BibitemOpen
  \bibfield  {author} {\bibinfo {author} {\bibfnamefont {Daniel}\ \bibnamefont
  {Foreman-Mackey}}, \bibinfo {author} {\bibfnamefont {David~W.}\ \bibnamefont
  {Hogg}}, \bibinfo {author} {\bibfnamefont {Dustin}\ \bibnamefont {Lang}}, \
  and\ \bibinfo {author} {\bibfnamefont {Jonathan}\ \bibnamefont {Goodman}},\
  }\bibfield  {title} {\enquote {\bibinfo {title} {{emcee: The MCMC Hammer}},}\
  }\href {\doibase 10.1086/670067} {\bibfield  {journal} {\bibinfo  {journal}
  {Publ. Astron. Soc. Pac.}\ }\textbf {\bibinfo {volume} {125}},\ \bibinfo
  {pages} {306--312} (\bibinfo {year} {2013})},\ \Eprint
  {http://arxiv.org/abs/1202.3665} {arXiv:1202.3665 [astro-ph.IM]} \BibitemShut
  {NoStop}%
\bibitem [{\citenamefont {{Goodman}}\ and\ \citenamefont
  {{Weare}}(2010)}]{2010CAMCS...5...65G}%
  \BibitemOpen
  \bibfield  {author} {\bibinfo {author} {\bibfnamefont {Jonathan}\
  \bibnamefont {{Goodman}}}\ and\ \bibinfo {author} {\bibfnamefont {Jonathan}\
  \bibnamefont {{Weare}}},\ }\bibfield  {title} {\enquote {\bibinfo {title}
  {{Ensemble samplers with affine invariance}},}\ }\href {\doibase
  10.2140/camcos.2010.5.65} {\bibfield  {journal} {\bibinfo  {journal}
  {Communications in Applied Mathematics and Computational Science}\ }\textbf
  {\bibinfo {volume} {5}},\ \bibinfo {pages} {65--80} (\bibinfo {year}
  {2010})}\BibitemShut {NoStop}%
\bibitem [{\citenamefont {Stanzione}\ \emph {et~al.}(2020)\citenamefont
  {Stanzione}, \citenamefont {West}, \citenamefont {Evans}, \citenamefont
  {Minyard}, \citenamefont {Ghattas},\ and\ \citenamefont
  {Panda}}]{10.1145/3311790.3396656}%
  \BibitemOpen
  \bibfield  {author} {\bibinfo {author} {\bibfnamefont {Dan}\ \bibnamefont
  {Stanzione}}, \bibinfo {author} {\bibfnamefont {John}\ \bibnamefont {West}},
  \bibinfo {author} {\bibfnamefont {R.~Todd}\ \bibnamefont {Evans}}, \bibinfo
  {author} {\bibfnamefont {Tommy}\ \bibnamefont {Minyard}}, \bibinfo {author}
  {\bibfnamefont {Omar}\ \bibnamefont {Ghattas}}, \ and\ \bibinfo {author}
  {\bibfnamefont {Dhabaleswar~K.}\ \bibnamefont {Panda}},\ }\bibfield  {title}
  {\enquote {\bibinfo {title} {Frontera: The evolution of leadership computing
  at the national science foundation},}\ }in\ \href {\doibase
  10.1145/3311790.3396656} {\emph {\bibinfo {booktitle} {Practice and
  Experience in Advanced Research Computing}}},\ \bibinfo {series and number}
  {PEARC ’20}\ (\bibinfo  {publisher} {Association for Computing Machinery},\
  \bibinfo {address} {New York, NY, USA},\ \bibinfo {year} {2020})\ p.\
  \bibinfo {pages} {106–111}\BibitemShut {NoStop}%
\end{thebibliography}%

\end{document}